\documentclass[floatfix,nofootinbib,12pt,prx]{revtex4}
\usepackage[utf8]{inputenc}
\usepackage{graphicx,amsfonts,amssymb,amsmath,hyperref,hypcap,enumerate}
\usepackage{slashed}
\usepackage{amsmath,amssymb,bm,graphicx,dsfont}
\usepackage{amsmath}
\usepackage{slashed}
\usepackage{dsfont}
\usepackage{amstext}
\usepackage{amssymb}
\usepackage{amsbsy}
\usepackage{amsthm}
\usepackage{epsfig}
\usepackage{graphicx}
\usepackage{braket}
\usepackage{bbm}
\usepackage{hyperref}
\usepackage{color}
\usepackage{xcolor}
\usepackage{ulem}
\usepackage{slashed}
\renewcommand{\v}[1]{\mathbf{#1}} 

\newcommand{\be}{\begin{equation}}
\newcommand{\ba}{\begin{align}}
\newcommand{\ee}{\end{equation}}
\newcommand{\bea}{\begin{eqnarray}}
\newcommand{\eea}{\end{eqnarray}}
\newcommand{\beq}{\begin{equation}}
\newcommand{\eeq}{\end{equation}}
\newcommand{\beqn}{\begin{eqnarray}}
\newcommand{\eeqn}{\end{eqnarray}}

\usepackage{graphicx}
\usepackage{dcolumn}
\usepackage{bm}

\renewcommand{\vec}[1]{{\bf #1}}
\renewcommand{\hat}[1]{{\widehat #1}}

\begin{document}
\title{Non-commutative field theory and composite Fermi Liquids in some quantum Hall systems}
\author{Zhihuan Dong}
\affiliation{Department of Physics, Massachusetts Institute of
Technology, Cambridge, MA 02139, USA}
 \author{T. Senthil}
\affiliation{Department of Physics, Massachusetts Institute of
Technology, Cambridge, MA 02139, USA}
\date{\today}

\preprint{APS/123-QED}

\begin{abstract}
Composite Fermi liquid metals arise at certain special filling fractions in the quantum Hall regime, and play an important role as parents of gapped states with quantized Hall response. They have been successfully described by the Halperin-Lee-Read (HLR) theory of a Fermi surface of composite fermions coupled to a $U(1)$ gauge field with a Chern-Simons term.  However  the validity of the HLR description when the microscopic system is restricted to a single Landau has not been clear. Here for the specific case of bosons at filling $\nu = 1$, we build on earlier work from the 1990s to formulate a low energy description that takes the form of a {\em non-commutative}  field theory. This theory has a Fermi surface of composite fermions coupled to a $U(1)$ gauge field with no Chern-Simons term but with the feature that all fields are defined in a non-commutative spacetime. An approximate mapping of the long wavelength, small amplitude gauge fluctuations yields a commutative effective field theory which, remarkably, takes the HLR form but with microscopic parameters correctly determined by the interaction strength. Extensions to some other composite fermi liquids, and to other related states of matter are discussed. 
    
\end{abstract}

\maketitle

\tableofcontents

\section{Introduction}
The celebrated quantum Hall effects occur when electrons move in two space dimensions in a large perpendicular magnetic field such that the  number of magnetic flux quanta  is comparable to the number of particles.  Our interest in this paper is in  metallic phases of matter - dubbed ``composite fermi liquids"(see Ref. \onlinecite{halperincflrev,willett97} for reviews) - that  play a foundational role in our overall understanding of phenomena in the quantum Hall regime. In electronic systems these occur at specific even denominator filling fractions $\nu = \frac{1}{2}, \frac{1}{4},.. $.  They are striking experimental examples of metals that are not standard Landau Fermi liquids. Further they act as ``parent phases" for an entire prominent sequence (the Jain states\cite{jainbook}) of  topological ordered states with quantized Hall conductivity that are 
observed experimentally.  

Theoretically the earliest and phenomenologically successful description of the experimentally observed metallic phase at $\nu = \frac{1}{2}$ was provided  by the seminal work\cite{HLR} of Halperin. Lee, and Read (HLR). The HLR theory employed a construction\cite{jaincf,lopez1991fractional} - known as flux attachment - where   the original interacting electron problem is formally rewritten in terms of a new fermionic degree of freedom (dubbed the composite fermion) together with a dynamical $U(1)$ gauge field with a Chern-Simons term.  In a mean field description, the composite fermion sees a reduced effective magnetic field $B^*$ compared to the physical magnetic field $B$. At filling $\nu = \frac{1}{2}$, the effective field $B^* = 0$, and the composite fermions form a Fermi surface. The HLR theory explained a number of essential experimental observations at $\nu = \frac{1}{2}$ and made further predictions that were confirmed in subsequent experiments.  The HLR theory also has received considerable numerical support.  This same general procedure was also extended  to $\nu = \frac{1}{4}$ and captured the observed physics. 

Despite its striking phenomenological success there were a number of fundamental theoretical questions and difficulties raised by the HLR theory that led to many further developments\ in subsequent years\cite{gmrsrmp03}. The most crucial  difficulty comes from considering the limit in which the Coulomb interaction energy is smaller than the Landau level spacing. This limit is routinely used in numerical calculations used to confirm the HLR theory, and  is not an unreasonable approximation for experiments. Then it is appropriate to define the quantum Hall problem by projecting the Coulomb interaction to the highest occupied Landau level and ignoring all other levels, (At $\nu = \frac{1}{2}$ this means that we define the problem entirely within the lowest Landau level.)  In this limit there is only interaction energy as the electron kinetic energy is completely quenched by the Landau level structure. Correspondingly there is a single energy scale that is set by the interaction strength; for the Coulomb interaction this energy scale is $\frac{e^2}{l_B}$ where $e$ is the electron charge and $l_B$ is the magnetic length. Formally the projection to the lowest Landau level can be thought of as the limit where the bare electron mass $m \rightarrow 0$. The HLR theory however is not faithful to this projection. Indeed in the mean field approximation of HLR the composite fermion effective mass is the same as the bare electron mass $m$.   Thus it is important to understand how to implement the physics of flux attachment while working purely within a single Landau level.  
A second crucial shortcoming - specific to electrons at $\nu = 1/2$ - has to do with a particle-hole symmetry present at that filling\cite{klkgphhlr} when the electron motion is restricted to a single Landau level.  This symmetry is known - through numerical calculations - to be preserved by the composite fermi liquid ground state\cite{rzyhald2000,geraedts2016half} but yet is not manifest in the HLR description and is possibly even absent.  This issue has attracted tremendous attention in recent years resulting in a proposal for a modified theory\cite{son2015composite} of the half-filled Landau level in terms of a Dirac composite fermion. This proposal has been substantiated through field theoretic duality transformations and other physical arguments\cite{wang2015dual,metlitski2016particle,wang2016half,mross2016explicit,karch2016particle,seiberg2016duality} and by numerical calculations\cite{geraedts2016half}. There have also been discussions of the emergence of particle-hole as an approximate symmetry within the HLR theory\cite{wang2017particle,kumar2019composite}. However all this progress on particle-hole symmetry did not address the basic issue of projecting to a single Landau level but rather sidestepped it (see, eg, the discussion in the review Ref. \onlinecite{senthil2019duality}). Hence we will not focus on it in this paper. 

Thus an outstanding question in the theory of composite fermi liquids  is to provide a microscopic derivation of an effective field theory by working within the Lowest Landau Level (LLL). Despite considerable theoretical attention\cite{rsgm97,Pasquier_1998,Read_1998,dhleephcf98,sternetal99} in the 1990s this  old question (for a review see Ref. \onlinecite{gmrsrmp03})  remained unanswered.    

Theoretically, composite fermi liquids are also expected to arise when the charged particles are bosons rather than fermions. For bosons at filling factor $\nu = 1$   attaching one flux quantum converts the bosons to composite fermions moving in an effective field $B^* = 0$. Then a composite Fermi liquid state can arise. Numerical calculations\cite{cooper2001quantum}  show that  the true ground state (with, say, a repulsive contact interaction) is an incompressible fractional quantum hall state - the bosonic Pfaffian - obtained by condensing pairs of composite fermions. Nevertheless the unpaired metallic composite fermion state is interesting to consider as a possible ground state (for some interaction), and as a `parent' state for understanding the bosonic Pfaffian. 
In pioneering work in the 1990s, Read developed\cite{Read_1998} a LLL theory of the composite fermi liquid state of bosons at filling $\nu = 1$. This work was based on an exact though redundant  representation\cite{Pasquier_1998}, introduced by Pasquier and Haldane, of the Hilbert space of these bosons in terms of composite fermions. The redundancy in the representation leads to constraints in the theory. A Hartree-Fock solution\cite{Pasquier_1998,Read_1998} leads to a compressible state with a composite Fermi surface. 
Fluctuations beyond Hartree-Fock were treated diagrammatically in Ref. \onlinecite{Read_1998} within a conserving approximation, and lead to physically sensible results for response functions similar to, but not identical to, those in the HLR theory. 

In this paper we will revisit the theory of Ref. \onlinecite{Read_1998} to pose and  answer several questions  that follow from it. What is a low energy  effective field theory for the composite Fermi liquid that results from this theory? How exactly is it related to the HLR action? Can one understand the emergence of the paired state in numerics within this microscopic analytic framework? Is it possible to generalize these results to other composite Fermi liquids? An answer to the first question was, in fact, suggested in Sec II.D of Read's paper. Specifically, the suggestion was that the low energy theory consists of a Fermi surface of composite fermions coupled to a dynamical $U(1)$ gauge field $a$ without a Chern-Simons term. The external background $U(1)$ gauge field $A$ was then proposed to couple linearly to $\frac{da}{2\pi}$.  This suggestion, which we review in Appendix \ref{app:priorftcfl} was however not explicitly obtained based on the diagrammatic calculations in the bulk of Ref. \onlinecite{Read_1998}.  A derivation of this suggested effective  theory using field theoretic duality transformations was also provided much later in Ref. \onlinecite{avl1} - however this derivation is not faithful to the LLL limit. Even later this effective action was also proposed\cite{wang2016composite} to arise when HLR is projected to the LLL limit through an emergent Berry phase of the composite fermions.  Assuming its correctness the Lagrangian of the suggested effective theory is clearly distinct from HLR; do the two Lagrangians describe the same universal aspects of the physics or do they describe distinct phases of matter?

In this paper we obtain a  low energy effective theory that captures the microscopic formulation and results of Ref. \onlinecite{Read_1998}. We show that this takes the form of a {\em non-commutative field theory}, {\em i.e} a theory defined in terms of fields that move in a space with non-commutative coordinates. The theory is expressed in terms of a single composite fermion field coupled to a non-commutative dynamical $U(1)$ gauge field $a$ without a Chern-Simons term. This is a precise formulation of the suggestion made in Ref. \onlinecite{Read_1998} but with many crucial differences. Specifically we will find that it does not have the form suggested there and which we review in Appendix \ref{app:priorftcfl}. We then use an approximate mapping - due to Seiberg and Witten\cite{Seiberg_1999} -  for the long wavelength, low amplitude gauge fluctuations between non-commutative and commutative field theories. We show that the resulting approximate commutative field theory action is precisely of the HLR form but with parameters  (like the composite fermion effective mass) faithful to the energetics of the LLL. 
We also examine the energetic stability of the paired state in a Hartree-Fock calculation within the Pasquier-Haldane-Read framework, and find that the paired state indeed wins in agreement with the numerics. Finally we will show that these methods can be readily generalized to describe composite fermi liquids formed by a system of spin-$1/2$ bosons at a total filling $\nu_T = 1$. Such a spinful bosonic composite fermi liquid has been found in numerical calculations.

Non-commutative field theories were first contemplated in physics a long time ago\cite{snyder}. Interest in them re-emerged in the late 1990s (and faded in the early 2000s) as they appeared in certain limits of string theory; for reviews see Refs. \onlinecite{douglas2001noncommutative,szabo2003quantum}.  
It has long been recognized that a single Landau level provides a wonderful physical example of non-commutative geometry (as the guiding center coordinates do not commute within a Landau level).  A rigorous proof of the quantization of the Hall conductivity in the integer quantum Hall effect used methods of non-commutative geometry\cite{bellissard1994noncommutative}.  For incompressible quantum Hall states a non-commutative effective field theory was proposed in Refs. \onlinecite{susskind2001quantum,polychronakos2001quantum}.  This is a possible  alternate  to the successful standard commutative Chern-Simons TQFT description of topologically ordered quantum Hall states, and  has the potential to be obtained from  a LLL description.  However we are not aware of such a microscopic derivation.  The non-commutative geometry of the Landau level also plays an important role in the work of Haldane\cite{haldane2011geometrical} on the  geometrical description of the fractional quantum Hall effect. For the composite fermi liquids of interest in this paper, the relevance of non-commutative geometry was pointed out  in Read's original paper\cite{Read_1998} but the full formulation in terms of non-commutative field theory was not developed.


\section{Preliminaries} 
\subsection{The basic problem}
\label{prlm_tbp}
We consider bosons of charge-$1$ in a magnetic field $B$ in two space dimensions at filling $\nu = 1$..  We take the bosons to occupy states in the LLL whose degeneracy we denote $N$.  Given a basis set $|m \rangle $ ($m = 1,....., N$) of one-particle states for the Landau level, the many particle Hilbert space is defined by the states 
\be
\label{bos_hs}
|\psi\rangle = \Sigma_{\{m_i\}} a_{m_1,......m_N} | m_1,........, m_N\rangle 
\ee
with the $a_{m_1,....,m_N}$ symmetric under permutations.  Clearly the number of particles equals the Landau level degeneracy reflecting the filling $\nu = 1$. 
The Hamiltonian is 
 \be 
 \label{bareH}
 {\cal H} = \frac{1}{2} \int \frac{d^2 \v q}{(2\pi)^2}  U(\v q)\rho_L(\v q) \rho_L(-\v q) 
 \ee
 Here $\v q$ is the momentum, and the Hermitian operators $\rho_L(\v q)$ satisfy the algebra\cite{GMP} 
 \be 
 \label{gmp}
 [\rho_L(\v q), \rho_L(\v q')] = 2i \sin\left( \frac{(\v q \times \v q') l_B^2 }{2} \right) \rho_L(\v q + \v q') 
 \ee
 This is known as the Girvin-MacDonald-Platzman (GMP) algebra. $l_B^2 = \frac{1}{B}$ is the magnetic length. The $\rho_L(\v q)$ are (up to an overall $\v q$ dependent factor that we absorb into the interaction) the physical density operators projected to the LLL.   Then we have 
 \be 
 U(\v q) =  e^{- \frac{q^2  l_B^2}{2}}U_0 (\v q)
 \ee
 where $U_0$ is the Fourier transform of the real space microscopic two-body repulsion between the bosons. We will work with a delta-function repulsion so that $U_0(\v q) = U_0$ independent of $\v q$. 
 Note that the only  length scale is $l_B$, and the only energy scale in the problem is $\frac{U_0}{l_B^2}$.  Unless specified we will henceforth work in units where  $l_B = 1$. 
 ( How this effective Hamiltonian is obtained by projecting a microscopic Hamiltonian with infinite number of Landau levels to the LLL is explained well in the literature, see e.g. Ref. \onlinecite{gmrsrmp03}.)

 We will begin with a representation of the GMP algebra in terms of canonical fermion operators $c_{\v k}$ found by Pasquier and Haldane\cite{Pasquier_1998}, and developed extensively by Read\cite{Read_1998}. We write 
 \be
 \label{phrrho}
 \rho_L(\v q) = \int \frac{d^2\v k}{(2\pi)^2}  c^\dagger_{\v k - \v q} c_{\v k} e^{i{\frac{  \v k \times \v q}{2}}}
 \ee
 The fermion operators satisfy the usual anticommutation relations 
 \be
 \{c_{\v k} , c^\dagger_{\v k'}\} = (2\pi)^2\delta^{(2)} (\v k - \v k') 
 \ee
 This is a redundant description, and requires dealing with a constraint 
 \be
 \label{phrcnstr}
  \rho_R (\v q) = \int  \frac{d^2\v k}{(2\pi)^2}  c^\dagger_{\v k - \v q} c_{\v k} e^{-i \frac{ \v k \times \v q}{2}}~= (2\pi)^2\underline{\rho} \delta^{(2)} (\v q) 
  \ee
 Here  $\underline{\rho} = \frac{B}{2\pi } = \frac{1}{2\pi l_B^2}$ is the mean density. 
  It is readily seen that $\rho_R$ satisfy a GMP algebra but with sign opposite to Eqn. \ref{gmp}. Furthermore $\rho_R$ commutes with $\rho_L$ at all momenta and hence with the Hamiltonian itself. The constraint operators may (for large but finite $N$) be thought of as generators of $U(N)$ gauge transformations (corresponding to a large redundancy in representing the physical Hilbert space in terms of the fermion operators).  The fermions $c_{\v k}$ are interpreted as (the LLL version of) the composite fermions. For a discussion of a physical picture in terms of vortex attachment to the particles, we refer to Ref. \onlinecite{Read_1998}.  Note that the $\v q \rightarrow 0$ limit of Eqn. \ref{phrrho} implies that the total number of composite fermions equals the number of physical bosons. 
  
Substituting Eqn. \ref{phrrho} into Eqn. \ref{bareH} gives a four-fermion Hamiltonian which must be solved together with the constraint Eqn. \ref{phrcnstr} imposed. 
A simple Hartree-Fock approximation that respects translation symmetry seeks a solution where 
\be 
\langle c^\dagger_{\v k} c_{\v k} \rangle \neq 0
\ee
The resulting Hartree-Fock Hamiltonian takes the form 
\be
{\cal H}_{HF} = \int \frac{d^2\v k}{(2\pi)^2} \epsilon_{\v k} c^\dagger_{\v k} c_{\v k} 
\ee
The composite fermions then form a Fermi sea, and we get a mean field description of a composite Fermi liquid.  To treat fluctuations beyond Hartree-Fock, we note that the Hartree-Fock ``order parameter" $c^\dagger_{\v k} c_{\v k} $ does not commute with 
$\rho_R(\v q)$ except at $\v q = 0$. Thus the huge group of gauge transformations generated by $\rho_R$ is broken spontaneously (Higgsed). The important fluctuations are those generated by  $\v q \approx 0$ - these can be thought of as a $U(1)$ gauge field. Thus we should expect to end up with an effective description in terms of a Fermi surface of composite fermions coupled to an emergent dynamical $U(1)$ gauge field.  The precise form of this effective theory and its relationship with HLR  will be discussed in the bulk of this paper. We will show that the effective theory is conveniently formulated as a non-commutative field theory and that HLR emerges in a long wavelength approximation. 

 \subsection{Non-commutative field theory} 
 \label{ncft}
 To set the stage we provide a lightning review of the basic formalism of non-commutative field theory. A detailed exposition may be found in Ref. \onlinecite{douglas2001noncommutative,szabo2003quantum}. Consider $2+1$-dimensional space-time where the two spatial coordinates $X$ and $Y$ do not commute: 
 \be
 [X, Y] = i \Theta 
 \ee
 Here $\Theta$ - known as the non-commutative parameter - is a constant. We can think of $X, Y$ as operators in a space of states. In the specific context of the Lowest Landau Level, $X,Y$ will be the components of the guiding center coordinate. They are operators in the space of single particle eigenstates of LLL. We are interested in fields that live in this non-commutative space. To that end first let us define scalar functions $f(\v R)$ (where $\v R = (X, Y)$). As $X, Y$ do not commute we need to specify what we mean by $f(\v R)$. A standard choice is known as the Weyl-ordering which defines functions in terms of their Fourier transform: 
 \be
 f(\v R) = \int \frac{d^2\v k}{(2\pi)^\frac{3}{2}} e^{i \v k \cdot \v R}\tilde{f}(\v k) 
 \ee
 Here $\tilde{f}(\v k)$ is an ordinary function of the ordinary momentum $\v k$ whose components commute with each other.  The ``plane wave" factor 
 \be
 \tau_{\v k} \equiv e^{i \v k \cdot \v R}
 \ee
 may be defined through its power series expansion and fixes the ordering of $X$ and $Y$.  The inverse Fourier transform is readily obtained: 
 \be
 \tilde{f}(\v k) = (2\pi)^{\frac{1}{2}}\int d^2{\v R} Tr \left(f(\v R) e^{- i \v k \cdot \v R}\right)
 \ee
 Here $\int Tr$ is over the space spanned by $\v R$. For notational convenience we will  drop the $Tr$ symbol in the subsequent equations and simply write $\int d^2\vec R$. 
 
 Note that as $\v R$ is an operator, we should  regard $f(\v R)$ also as an operator. It will be convenient to associate to this operator an ordinary function $f(\v x)$ of commuting coordinates $\v x$ by taking the ordinary inverse transformation of $\tilde{f}(\v k)$. 
 
 Next consider the product of two operator-valued functions $f(\v R)$ and $g(\v R)$. It is easy to see from the definition through the Fourier transforms that 
 \be 
 \label{op_p}
 f(\v R)  g(\v R) = \int \frac{d^2\v k d^2 \v k'}{(2\pi)^3} \left( e^{-i \frac{\Theta \v k \times \v k'}{2}}\tilde{f}(\v k) \tilde{g} (\v k')\right) e^{i (\v k + \v k') \cdot \v R}
 \ee
 Taking the ordinary inverse Fourier transform, we find that the ordinary function $f(\v x) * g(\v x)$ that corresponds to this product is not the ordinary product of functions but rather to a modification known as the star product. 
 Thus the Fourier transform of $f(\v x) *g(\v x)$ at momentum $\v q$ is 
 \be
 \label{starp}
 \int \frac{d^2\v k}{(2\pi)^\frac{3}{2}} \tilde{f}(\v q - \v k) \tilde{g}(\v k) e^{-i \frac{\Theta \v k \times \v q}{2}}
 \ee
 The  star product is associative but not commutative. We will henceforth remove the tilde from the momentum space variables and simply use the argument (coordinate versus momentum space) as an identification of which object we are talking about.  Derivatives of operators  can also be readily defined and correspond to ordinary derivatives of the functions $f(\v x)$. 
 
 

Given an operator-valued field $\phi(\v R)$, consider a term in a putative Lagrangian such as $\phi(\v R) \phi(\v R)$. It's integral $\int d^2\v R \phi(\v R)  \phi(\v R)$ can be written as an ordinary Fourier space integral and hence is precisely defined. Thus we can build field theories defined in non-commutative space. Equivalently we can also  work with the corresponding ordinary fields $\phi(\v x)$ and write the action with all products being star-products. For instance the non-commutative $\phi^4$ theory has an action that can be written 
\be
S = \int d\tau d^2\v x  \left[\partial_\mu \phi * \partial_\mu \phi + r \phi * \phi + u \phi * \phi * \phi * \phi \right]
\ee
From the definitions above we see that the non-commutativity shows up only in the product that defines the quartic term. 

Non-commutative gauge theories can be similarly defined. We will only need to work with $U(1)$ gauge fields $a_\mu(\v R, \tau)$ which will again be defined in terms of their Fourier transform, or the corresponding $a_\mu(\v x, \tau)$. We use the latter below.  The corresponding field strength is
\be
f_{\mu \nu } = \partial_\mu a_\nu - \partial_\nu a_\mu + i [a_\nu, a_\nu ]_*
\ee
Here we introduced the $*$-commutator
\be
[A, B]_* = A*B - B*A
\ee
Gauge transformations correspond to 
\be
a_\mu \rightarrow a_\mu + \partial_\mu \lambda + i[a_\mu, \lambda]_*
\ee
These leave the field strength $f_{\mu \nu}$ invariant.

\subsection{Summary of results}
We can now state the main results of this paper. We begin by showing that an effective low energy theory that describes the results of Ref. \onlinecite{Read_1998} is a non-commutative gauge theory with the imaginary time Lagrangian in terms of a composite fermion field $c$:
\be
\label{nccfllag}
{\cal L} = \overline{c} * D_0 c + i a_0  \underline \rho +  \frac{1}{2m^*} \overline{D_i c} D_i c \ee
Here the covariant derivatives are defined through 
\be
\label{nccdrv}
D_\mu c = \partial_\mu c - i c*a_\mu - i A_\mu* c
\ee
where $a_\mu$ ($\mu = 0, 1,2$) is a dynamical $U(1)$ gauge field and $A_\mu$ is an external background $U(1)$ gauge field. The non-commutative parameter $\Theta = - l_B^2$.  The composite fermion effective mass $m^*$ is determined by the interaction strength. The composite fermions have a density $\underline{\rho}$. The theory thus takes the expected form of a Fermi surface of composite fermions coupled to a dynamical $U(1)$ gauge field. 
Furthermore, in agreement with what is claimed  in Ref. \onlinecite{Read_1998}, this theory has no Chern-Simons term for the dynamical gauge field. 

However these results come with the added feature not mentioned in Ref. \onlinecite{Read_1998} - namely this is a non-commutative field theory. In contrast the standard descriptions (such as HLR)  of composite fermi liquids obtained without paying restricting to the LLL is in terms of commutative field theories. We will show that there is an approximate mapping of the theory in Eqn. \ref{nccfllag} to a commutative field theory which remarkably is the same as the effective action of HLR (with some calculable subleading corrections) but with an effective mass set by the interaction strength. 

The key technical tool we use is known as the Seiberg-Witten map. This map enables trading a non-commutative field theory for a commutative one in a systematic  series expansion
\footnote{For pure non-commutative $U(1)$ gauge theory, there is an exact non-perturbative version\cite{liu2001trek,okawa2001exact,mukhi2001gauge,liu2001ramond} of the Seiberg-Witten map relating it to a commutative $U(1)$ gauge theory. A physically appealing understanding\cite{jackiw2002noncommuting} (see also Ref. \cite{susskind2001quantum}) of this result relates it to the map in fluid dynamics between the Lagrangian and Eulerian frameworks.}
in powers of the non-commutative parameter $\Theta = - l_B^2$. As $\Theta$ is dimensionful this should really be regarded as an expansion in $(l_Bq)^2,  l_B^2\delta \rho_L$ where $q$ is the momentum of the gauge field, and $\delta \rho_L$ is the deviation of the real space density from it's mean, {\em i.e} as a  long wavelength, low amplitude,  expansion. 
 
 Thus we conclude that though the non-commutative field theory Eqn. \ref{nccfllag} is a more microscopically faithful effective theory, its approximate equivalence to HLR in the long wavelength limit vindicates the use of HLR for addressing many universal physical properties. 
 
 We also examine the energetics of pairing of composite fermions within the Hartree-Fock theory. Within this treatment we find that the composite Fermi liquid is unstable to pairing.  However the pairing gap is numerically small compared to the fermi energy of the composite fermions. Thus fluctuations beyond mean field may affect the relative stability of the paired state as compared to the composite fermi liquid. In particular it is known that gauge fluctuations oppose pairing. Nevertheless the mean field stability of the paired state is encouraging and agrees with existing numerical results. 
 
 We generalize these methods to study the problem of 2-component bosons with full $U(2)$ symmetry at a total filling fraction $\nu_T = 1$. Numerical work\cite{wu2015emergent,geraedts2017emergent} has indicated the presence of a spin unpolarized composite fermi liquid at this filling. We  develop an effective non-commutative field theory of this composite Fermi liquid, and show again that it reduces to a HLR form in the long wavelength, low amplitude approximation. At the mean field level we find a pairing instability which is somewhat weaker than for the spinless case.

\section{Pasquier-Haldane-Read construction for composite fermion}
\label{sec:PH}
\subsection{Parton construction and gauge structure}
\label{sec:parton construction}

We begin with a brief discussion of some physical pictures\cite{read94} that motivate the formal parton construction of the composite fermion. 
In the LLL, the process of flux attachment to the particles to produce the composite fermion is replaced by the concept of ``vortex attachment".  The vortex comes with a depletion of charge density at its core. A heuristic argument shows that the depletion is precisely equal to the boson charge. Thus we may view the composite fermion as a bound state of the boson with electric charge $+1$ and a vortex with electric charge $-1$. Such a bound state is electrically neutral but has a dipole moment $\v d$  that is determined\footnote{This follows from the relation between position and momentum in the LLL.} by the composite fermion momentum $\v k$: 
\be
\label{dipole}
\v d = l_B^2  \v k \times \hat{z}
\ee
This dipole moment gives an appealing physical picture for how the composite fermion gets a dispersion. It is simply the polarization energy $\sim \vec d^2$ (for small $|\vec d|$) which leads to a $\vec k$-dependent energy. 

The Pasquier-Haldane formalism begins with a redundant description of the Hilbert space of bosons at $\nu = 1$ in terms of the composite fermion Hilbert space. We will follow the presentation in Ref. \onlinecite{Read_1998}. Introduce the fermion operators $c_{nm}, c^\dagger_{mn}$ satisfying 
\be
\{c_{mn}, c^\dagger_{m'n'} \} = \delta_{mm'} \delta_{nn'}
\ee
Here $m, n$ are integers that range from $1$ to $N$. Below we will identify $N$ to be the total number of  orbitals in a single Landau level, and will eventually take $N \rightarrow \infty$. The basis states of the physical Hilbert space (see Eqn. \ref{bos_hs}) of the bosons is then constructed as 
\be
|m_1,......,m_N \rangle = \epsilon^{n_1 n_2.....n_N}c^\dagger_{n_1m_1} c^\dagger_{n_2m_2}......c^\dagger_{n_Nm_N}|0\rangle
\ee
Here $|0\rangle$ is the Fock vacuum annihilated by all the $c$-operators, and $\epsilon$ is fully antisymmetric with $\epsilon^{12....N} = 1$. A sum over repeated indices is assumed. Clearly the states thus constructed are fully symmetric (and hence describe bosons), and correspond to a total number $N$ of bosons in the $N$ orbitals. Thus we describe bosons at filling $\nu = 1$. The antisymmetrization implied by the $\epsilon$ symbol implies that physical states are singlets under $SU(N)$ transformations  of the $n$ index of the $c_{mn}$ operators. The generators of these transformations are 
\be
\label{cnstrntmtrx}
\rho^R_{nn'} = c^\dagger_{nm}c_{mn'} 
\ee
Here we have included a global $U(1)$ generator given by $c^\dagger_{nm}c_{nm}$ so that the $\rho^R$ generate $U(N)$ transformations. The constraint that physical states are $SU(N)$ singlets can be restated in terms of these operators as 
\be
\rho^R_{nn'} |\psi_{phys} \rangle = \delta_{nn'} |\psi_{phys} \rangle 
\ee
These $SU(N)$ transformations express an $SU(N)$ gauge redundancy in the representation of the boson Hilbert space in terms of fermions. 
We can similarly define unitary $U(N)$ transformations on the physical `left' index $m$ generated by 
\be
\rho^L_{mm'} = c^\dagger_{nm'}c_{mn} 
\ee
Unlike the transformations on the right index, the transformations generated by $\rho^L$ are physical operations in the boson Hilbert space. Note that the overall $U(1)$ generator $N = Tr \rho^L = Tr \rho^R$ is shared by both the left and right generators (and in any case is fixed to be $N$ as we work with a  fixed total number of bosons). 
We will refer to $\rho^L$ and $\rho^R$ as the left and right densities respectively. 

We will now proceed slightly differently from the development in Ref. \onlinecite{Read_1998} in a manner suitable for our purposes. We will identify the $m$ and $n$ indices to denote orbitals in a single Landau level. Then we may regard the matrix $c_{nm}$ as the matrix elements of an abstract operator $c$ defined in such a Landau level. 
As is well known a single Landau level is a non-commutative space. Specifically consider the guiding center coordinate $\v R$ of a single  particle moving in the LLL. It's components $X, Y$ satisfy 
\be
[X, Y] = -i l_B^2
\ee
We can use $\v  R$ to define a magnetic translation operator 
\be
\tau_{\v k} = e^{i \v k \cdot \v R}
\ee
Any matrix in the LLL can be expanded in terms of the matrix elements of $\tau_{\v k}$. Thus we write 
\be
c_{mn} = \int \frac{d^2\v k}{(2\pi)^2} \langle m | \tau_{\v k} |n \rangle c_{\v k} 
\ee
The momentum space operator $c_{\v k}$ is readily seen to satisfy the usual anti-commutation relations
\be 
\{ c_{\v k}, c^\dagger_{\v k'} \} = (2\pi)^2\delta^{(2)} (\v k - \v k') 
\ee
 The densities $\rho^{L,R}$ may also be similarly expressed in momentum space. We define
 \bea
 \rho^L_{mm'} & = &  \int \frac{d^2\v k}{(2\pi)^2} \langle m |\tau_k |m'\rangle \rho^L({\v k}) \\
\rho^R_{nn'} & = &  \int \frac{d^2\v k}{(2\pi)^2} \langle n |\tau_k |n'\rangle \rho^R({\v k} )
\eea

It is readily seen that $\rho^L(\v k)$ satisfies the standard GMP algebra of Eqn.~\ref{gmp} as expected for the physical boson density. The $\rho^R(\v k)$ also satisfies a GMP algebra but with the opposite sign: 
\be 
 \label{gmpR}
 [\rho^R(\v q), \rho^R(\v q')] = - 2i \sin\left( \frac{(\v q \times \v q') l_B^2 }{2} \right) \rho^R(\v q + \v q') 
 \ee
 Furthermore $\rho^L$ and $\rho^R$ commute with each other.  In momentum space, the constraint in Eqn. \ref{cnstrntmtrx} then becomes exactly Eqn. \ref{phrcnstr}. Furthermore the left and right densities - when expressed in terms of the momentum space fermion operators - take precisely the forms given in Eqns. \ref{phrrho} and \ref{phrcnstr}. Thus as advertised before, the momentum space version of the Pasquier-Haldane representation faithfully reproduces the Hilbert space of bosons at $\nu = 1$. 
 
 The $c(\v k)$ are identified with composite fermion destruction operators at momentum $\v k$. 
To bolster this interpretation, let us establish that these fermions have the correct dipole moment given by Eqn. \ref{dipole}. 

\subsection{Dipole moment of the composite fermion} 
Consider the deviation $\delta \rho_L(\v q)$ of the physical density operator  from it's mean:
\be
\delta \rho_L(\v q) = \rho_L(\v q) - \rho_L(\v q = 0) 
\ee
Expanding to $o(\v q)$ we get
\be
\label{drlim}
\delta \rho_L(\v q) \simeq \int \frac{d^2k}{(2\pi)^2} \left(  \frac{i}{2} \v k \times \v q c^\dagger_{\v k} c_{\v k} - \v q \cdot \frac{\partial c^\dagger_{\v k}}{\partial \v k} c_{\v k} \right) + o(q^2) 
\ee

The second term involving $\frac{\partial c^\dagger_{\v k}}{\partial \v k} c_{\v k}$ can be simplified using the definition of $c_{\v k}$ but we will use a different argument to obtain its form in the physical Hilbert space. To the same order in $\v q$, the deviation $\delta \rho_R(\v q)$ of the constraint density is 
\be
\delta \rho_R(\v q) \simeq \frac{d^2k}{(2\pi)^2} \left( -\frac{i}{2} \v k \times \v q  c^\dagger_{\v k} c_{\v k} - \v q \cdot \frac{\partial c^\dagger_{\v k}}{\partial \v k} c_{\v k} \right) + o(q^2) 
\ee

Acting on physical states in the Hilbert space  we must have 
\be
\label{gc}
\delta \rho_R(\v q)|\psi_{phys} \rangle  = 0 
\ee
It follows that 
\be
\int \frac{d^2k}{(2\pi)^2}  \v q \cdot \frac{\partial c^\dagger_{\v k}}{\partial \v k} c_{\v k} |\psi_{phys}\rangle
= - i \int \frac{d^2k}{(2\pi)^2}   \frac{\v k \times \v q}{2} c^\dagger_{\v k} c_{\v k}|\psi_{phys} \rangle
\ee
Substituting in Eqn. \ref{drlim} we get 
\be
\delta \rho_L(\v q)|\psi_{phys} \rangle \simeq i\int  \frac{d^2k}{(2\pi)^2} ~\v k \times \v q c^\dagger_{\v k} c_{\v k}|\psi_{phys} \rangle
\ee

In the long wavelength limit, in real space, we write 
\be 
\delta \rho_L (\v x) = - \v \nabla \cdot \v P
\ee
where $\v P$ is the net dipole moment per unit area. In Fourier space we then have $\delta \rho_L(\v q) =  - i \v q \cdot \v P_{\v q} \simeq -i \v q \cdot P_{\v q = 0}$ to leading order in $\v q$. Thus we identify 
\be
\v P_{\v q = 0} = \int  \frac{d^2k}{(2\pi)^2} ~ \v k \times \v{\hat{z}} c^\dagger_{\v k} c_{\v k}
\ee
Thus each fermion at momentum $\v k$ can be assigned a dipole moment exactly as given by Eqn. \ref{dipole}. 

\subsection{Hartree-Fock theory} 
Substituting the expression for $\rho_L$ in terms of the $c$-fermions leads to a four-fermion Hamiltonian. In a Hartree-Fock treatment Ref. \onlinecite{Read_1998} sought for and found a ground state with a filled Fermi sea. Here we will modify this treatment in two different ways. First rather than write the Hamiltonian just in terms of $\delta \rho_L$ we will use $\rho_L - \rho_R$ for the (deviation from the mean of the) physical density.  A second modification is that we will allow for mean field states where the composite fermions are paired. 

We start with the Hamiltonian
\begin{equation}
\label{eq:Hamiltonian}
    H = \frac{1}{2}\int \frac{d^2q}{(2\pi)^2} \tilde{U}(\v q):\rho^L_{\v q}\rho^L_{-\v q}:
\end{equation}
together with the gauge  constraint
\begin{equation}
    \rho^R_{\v q}\ket{\psi_{phys}}= \underline{\rho} (2\pi)^2 \delta^2(\v q)\ket{\psi_{phys}}
\end{equation}
The normal ordering in Eqn. \ref{eq:Hamiltonian} is potentially important when we use the expression Eqn. \ref{phrrho} for $\rho_L$ in terms of the composite fermion operators. However it is readily checked that  
\be
:\rho_L(\v q) \rho_L(- \v q): - \rho_L(\v q) \rho_L(-\v q) =  - \int \frac{d^2k}{(2\pi)^2}\: c^{\dagger}_{\v k}c_{\v k}~~= -\underline{\rho}
\ee
As this is just a constant we can remove the normal ordering from Eqn. \ref{eq:Hamiltonian}. 
We specialize to a $\delta$-function repulsive interaction, with the projected two-body potential $\tilde{U}(\v q)=U(\v q)e^{-q^2/2}=Ue^{-q^2/2}$. 

It is straightforward to insert Eqn. \ref{phrrho} and do a Hartree-Fock mean field of the resulting Hamiltonian, as was done by Read\cite{Read_1998}. Restricting to unpaired translation-invariant states, the composite fermion acquires a dispersion. However, as observed already in Ref. \onlinecite{Read_1998} this mean field treatment  has some physically unsatisfactory features. To see this note that the dispersion in this mean field is given by
\be
\label{eq:readdisf}
\epsilon_k = \frac{1}{2}\tilde{U}(0) \int \frac{d^2 k'}{2\pi} \langle c^\dagger_{\v k'} c_{\v k' }\rangle  - \frac{1}{2}\int \frac{d^2k'}{2\pi} \tilde{U}(\v k - \v k') \langle c^\dagger_{\v k} c_{\v k'} \rangle
\ee
The first is the Hartree-term and the second is the Fock term. Thus the dispersion comes entirely from the Fock term. At first sight this gives a sensible dispersion $\epsilon_k$ which increases monotonically as $k$ increases. However the origin of the dispersion through the Fock term  is different from  the physical picture for how the composite fermion gets a dispersion, namely from  the polarization energy due to it's 
dipole moment. This polarization energy should have been a Hartree effect which, within this mean field, does not affect the dispersion. 

This problem appears in a much more severe form if we we were to use the same formalism to treat a system of fermions at $\nu = 1$. Then instead of composite fermions $c_{\v k}$ we would introduce composite bosons $b_{\v k}$. Proceeding as above would give a boson dispersion of the same general structure as Eqn. \ref{eq:readdisf} except that the Fock term now comes with a positive sign. This yields a  problematic composite boson dispersion that monotonically {\em decreases} with $k$. Indeed in this problem we simply expect to condense the composite boson at $\v k = 0$ so as to recover the obvious answer (of a fully filled Landau level with an Integer Quantum Hall effect). The inability of this mean field to capture this physics suggests that it is not a good physical starting point.

Here  we will rectify these problems by  modifying the Hamiltonian (following Ref. \onlinecite{murthy2016nu}) by replacing $\delta \rho_L$ by $\rho_L - \rho_R$. Within the physical Hilbert space both these expressions  have the same matrix elements. However as seen in the previous subsection, $\rho_L - \rho_R$ has the right long wavelength dipole moment and hence leads to a more physical Hartree-Fock mean field Hamiltonian.

 Thus we work with the modified Hamiltonian 
 \be
 \label{eq:modHamiltonian}
    \tilde{H}  = \frac{1}{2}\int \frac{d^2q}{(2\pi)^2} \tilde{U}(\v q)\left( \rho^L_{\v q}- \rho^R_{\v q} \right) \left( \rho^L_{-\v q} - \rho^R_{-\v q}\right) 
\end{equation}
We begin by writing this in normal ordered form: 
\be
\label{eq:Hamiltonian for HFMF}
\begin{split}
 \tilde {H} & =  \int \frac{d^2k}{2\pi}\: \int \frac{d^2q}{2\pi}\: 2\tilde{U}(\v q)\sin^2{\left(\frac{\v k\times \v q}{2}\right)}c^{\dagger}_{\v k}c_{\v k} + \frac{1}{2}
    \int \frac{d^2q}{(2\pi)^2}\:
    \tilde{U}(\v q):(\rho^L_{\v q}-\rho^R_{\v q})(\rho^L_{-\v q}-\rho^R_{-\v q}): \\
& =  \int \frac{d^2k}{2\pi}\: (1-e^{-\frac{k^2}{2}})U c^{\dagger}_{\v k}c_{\v k} +
    \frac{1}{2}\int \frac{d^2q}{(2\pi)^2}\:
    \tilde{U}(\v q):(\rho^L_{\v q}-\rho^R_{\v q})(\rho^L_{-\v q}-\rho^R_{-\v q}):
\end{split}
\ee

The two terms in the last line of Eqn. \ref{eq:Hamiltonian for HFMF} have clear physical meaning. The first term is a one-body term  for the bare self energy of a single dipole. This is precisely the polarization energy of the dipole which should contribute to the dispersion.  Despite being interaction-driven it  does not depend on the occupation $n_k$ of other dipole states.   The second  term can be thought of as a dipole-dipole interaction, which is still a two-body term that needs a mean-field treatment. 

\begin{figure}
    \centering
    \includegraphics[width=0.48 \textwidth]{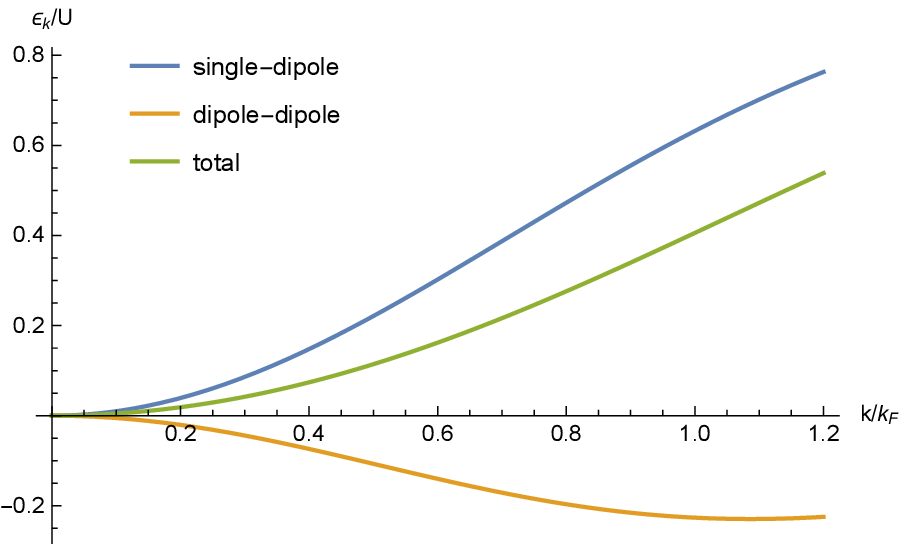}
    \includegraphics[width=0.48 \textwidth]{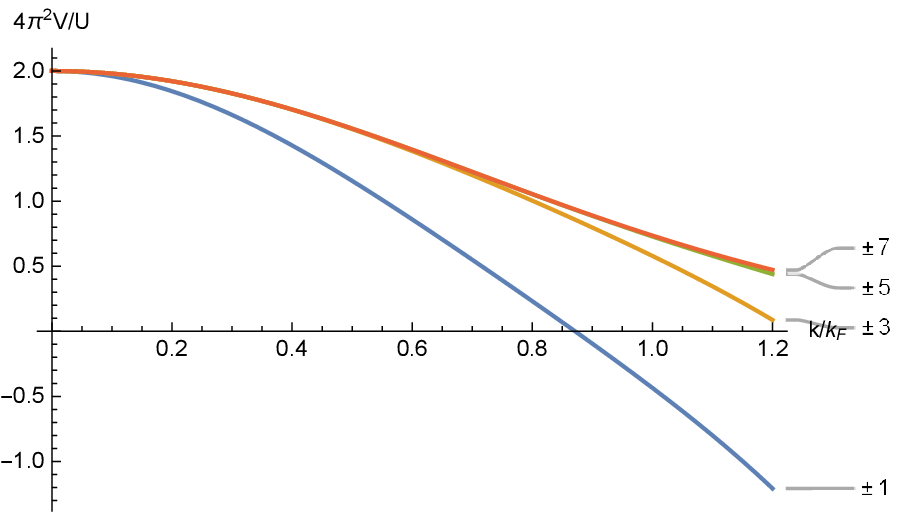}
    \caption{Mean-field dispersion and partial wave components for different angular momentum. Horizontal axis show $k/k_F$. (a) Composite fermion dispersion at mean-field level without pairing included. The blue and orange curves denote contributions from normal-ordering (marked as ``single-dipole") and from normal-ordered term (marked as ``dipole-dipole"), respectively. The green curve, their sum, is the total dispersion.  (b) Partial wave components for different pairing channels, as in Eqn. \ref{eq:partial wave}. The labels stand for corresponding angular momentum $l$. At the Fermi surface of composite fermion, only $l=\pm 1$ is attractive.}
    \label{fig:mean-field dispersion&pairing}
\end{figure}

The Hartree-Fock calculation procedure is standard. For the quartic term in Eqn.~\ref{eq:Hamiltonian for HFMF}, we adopt the translation and rotational symmetric mean-field ansatz
\be
\begin{split}
    \langle c^{\dagger}_{\v k} c_{\v k'}\rangle &= \delta^{(2)}(\v k-\v k')n_k\\
    \langle c_{\v k} c_{\v k'}\rangle &= \delta^{(2)}(\v k+\v k')d(k) = \delta(\v k+\v k')e^{il\theta_k}d(|k|)\\
\end{split}
\ee
where the pairing angular momentum $l$ is an odd integer. The effective mean field  Hamiltonian is reduced into a quadratic form
\be
    H_{MF} =\int \frac{d^2k}{2\pi}\:  \epsilon_k c^{\dagger}_{\v k} c_{\v k} + \Delta_{\v k} c_{\v k} c_{-\v k} + \Delta_{\v k}^* c^{\dagger}_{-\v k} c^{\dagger}_{\v k}
\ee
We begin by ignoring the possibility of pairing. The best unpaired state has a filled circular Fermi surface (centered at $\v k = 0$) of composite fermions with a Fermi momentum $k_F$ determined in the usual way: 
\be
\frac{\pi k_F^2}{(2\pi)^2} = \underline{\rho}
\ee
We identify this state with (the mean field description of) the composite fermi liquid. 
The composite fermion dispersion $\tilde{\epsilon}_k$, with occupation $n_{ k}$ given by the Fermi sea, is
\be
\label{eq:HF dipersion}
    \tilde{\epsilon}_{\v k}=U \left(1-e^{-\frac{k^2}{2}}\right) - 2 U e^{-\frac{k^2}{2}}\int_0^{k_F} \frac{dk'}{2\pi}\: k' e^{-\frac{k'^2}{2}}\left(I_0\left(kk'\right)-1\right)
\ee
where $I_l(z)$ is the modified Bessel function. This modified dispersion replaces the one above in Eqn. \ref{eq:readdisf}. The first term arises from the polarization energy of a single dipole and the second term from the dipole-dipole interaction. In Fig.~\ref{fig:mean-field dispersion&pairing}, where the contributions from two terms are plotted separately, we show that the self-energy term dominates over dipole-dipole term for all $k$, which is sensible, since the inter-dipole contribution only screens and weakens the intra-dipole interaction. It can be checked that $\tilde{\epsilon}_k$ is a monotonically increasing function of $k$. 
The composite fermion effective mass $m^*$ for states near the Fermi surface is given (within the Hartree-Fock approximation), as usual, by 
\be
\label{meff}
\frac{K_F}{m^*} = \left[\frac{\partial \tilde{\epsilon}_{\v k}}{\partial k}\right]_{k = K_F}
\ee
Numerically we find the value of effective mass to be $m^*=\frac{1.54}{U_0}$.

If instead we had solved the problem of fermions at $\nu = 1$ within the same framework using composite bosons, the coefficient of the first term in Eqn. \ref{eq:HF dipersion} is unaffected  while the second term has the opposite sign.   As the first term dominates the  composite boson dispersion   has its  minimum at $k=0$, and increases with increasing $k$. Thus this modified mean field yields physically sensible answers for both bosons at $\nu = 1$ and for fermions at $\nu = 1$. 

Returning to bosons at $\nu = 1$, the ground state energy of the mean field composite fermi liquid is $0.2913UN$, where $N$ is the Landau level degeneracy. 

We now include the possibility of pairing of composite fermions. We will assume that any pairing that is found is weak in the sense that the pairing gap $\Delta$ is small compared to the Fermi energy $E_F$ of the composite fermions. Note that as the only energy scale in the problem is $U$, there can be no parametric separation between $\Delta$ and $E_F$. Nevertheless this assumption is justified a posteriori as the solution we will find will have small $\frac{\Delta}{E_F}$.

First we  show that $l=\pm 1$ is the only possible pairing channel by examining the partial wave components of the pairing potential
\be
\label{eq:partial wave}
\begin{split}
    V_l(k)&= \frac{1}{2} U e^{-k^2}\int_0^{2\pi} \frac{d\theta}{2\pi} \left(e^{i\frac{k^2}{2}\sin{\theta}}-c e^{-i\frac{k^2}{2}\sin{\theta}}\right)^2 e^{k^2\cos{\theta}}e^{-i l\theta}\\
    &= -U e^{-k^2} \left(I_l\left(2k^2\right)-1\right)
\end{split}
\ee
The behavior of $V_l(k)$ is plotted in Fig. \ref{fig:mean-field dispersion&pairing}. At the Fermi surface, only $l=\pm 1$ channels are attractive.

\begin{figure}
    \centering
    \includegraphics[width= \textwidth]{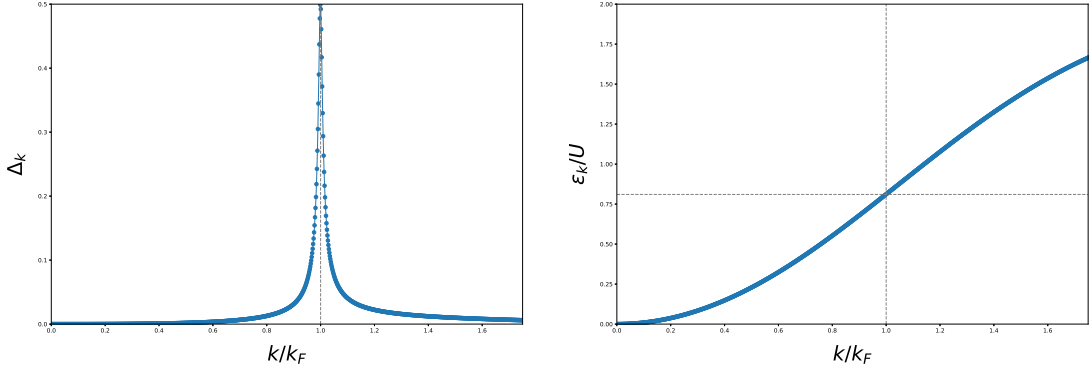}
    \caption{The self-consistent Hatree-Fock mean-field solution without single particle potential $V$, and with $l=+1$ pairing channel turned on. (a) $p+ip$ pairing order parameter. (b) Dispersion of composite fermion induced by interaction. Horizontal dotted line shows the  
    chemical potential and the vertical dotted line $k = k_F$.}
    \label{fig:self-consistent mean-field}
\end{figure}

Now we  carry out a numerical self-consistent calculation. As shown in Fig.~\ref{fig:self-consistent mean-field}, we find a stable solution in the $l=\pm1$ pairing channel, whose energy is $0.2908U$, below that of the composite Fermi liquid solution. Furthermore  the pairing is weak ($\Delta/E_F \ll 1$) thereby justifying our approximations.   

In numerical exact diagonalization calculations\cite{cooper2001quantum} of the microscopic Hamiltonian for bosons at $\nu = 1$ it is found that the ground state is a gapped topologically ordered state that can be thought of as a paired state of composite fermions, and not the composite fermi liquid. It is interesting that the mean field treatment described here captures this preference for a paired state. However this mean field treatment also has an artifact, namely the degeneracy between $l=\pm 1$ pairing channels. This is a direct consequence of identifying the physical density operator with $\rho^L-\rho^R$, which makes the interaction term symmetric\cite{murthy2016nu} under a  discrete anti-unitary operation that interchanges $\rho_L$ and $\rho_R$. The constraint, however, does not have this symmetry. Thus we expect that fluctuations beyond the mean field will select between these two possibilities. The two mean-field paired ground states correspond to two topologically distinct states, both of which has a charged edge mode propagating along the direction fixed by the Landau level, and a neutral edge mode propagating in parallel or anti-parallel directions.  

Though the paired state wins over the composite fermi liquid in the mean field, it is still interesting to consider the fate of the composite fermi liquid beyond mean field. The energetic preference for the paired state may be altered by a different microscopic interaction; furthermore it is conceivable, as the pairing is in any case weak,  that there is a temperature window in which the physics of the composite fermi liquid is relevant even if the true ground state is paired. 


\section{Fluctuations about the mean field composite fermi liquid: Non-commutative field theory}
\label{sec4}
In this section we will go beyond the mean field treatment and incorporate fluctuations to obtain an effective action for the  composite fermi liquid. 

The mean field CFL state is not invariant under the `right' gauge transformations generated by $\rho_R(\v q)$ (except in the trivial limit $\v q = 0$ which does not correspond to a generator of a gauge transformation): 
\be
\label{rightg}
c_{mn} \rightarrow c_{mn'} U^R_{n'n}
\ee
where $U^R$ is an $SU(N)$ matrix.  As the $q = 0$ $U_R$ rotations are unbroken, the important fluctuations beyond mean field are long wavelength rotations by $U_R$.  To capture these  we will introduce a dynamical   $U(1)$ gauge field $a_\mu$ that couples to the `right' 3-currents of the $c$ fermions.  We will also include a coupling to a background non-commutative gauge field $A_\mu$  that corresponds to `left' gauge transformations generated by $\rho_L$: 
\be
\label{leftg}
c_{mn} \rightarrow U^L_{mm'}c_{m'n}
\ee
with $U^L$ another $SU(N)$ matrix. 

We begin with the effective Hartree-Fock action for the composite fermions. In what follows we will replace  the Hartree-Fock dispersion $\epsilon_{\v k}$ by a simpler quadratic dispersion: 
\be
\tilde{\epsilon}_{\v k} \rightarrow \frac{\v k^2}{2m^*}
\ee
with the $m^*$ given in Eqn. \ref{meff}. 
As the low energy physics is dominated by states near the Fermi surface anyway this replacement is innocuous: it only modifies the dispersion away from the Fermi surface. The imaginary time Hartree-Fock action may then be written
\be
{\cal S}_{HF} = \int d\tau \frac{d^2\v k}{(2\pi)^2} ~~\overline{c}_{\v k} \frac{d c_{\v k}}{d\tau} - \left(\frac{\v k^2}{2m^*}   \right) c^\dagger_{\v k} c_{\v k} 
\ee
It is understood that the fermions are at a non-zero mean density $\underline{\rho}$. This could be implemented explicitly by including a chemical potential term but we will not do so. 

To proceed we could try to continue to work in $k$-space; however the action of the gauge fluctuations in $k$-space is complicated and mixes 
fermion operators at different momenta. We could try going back to the matrix operators $c_{mn}$ on which the gauge transformations act simply. 
However then the mean field action looks complicated\footnote{The same is true if we choose explicit wavefunctions for the 
Landau orbitals, for instance, in the symmetric gauge to define fermion operators\cite{Read_1998} $c(z, \bar{w})$ as a function of a holomorphic coordinate $z$ and an anti-holomorphic coordinate ${\bar{w}}$.}.  

These problems are nicely circumvented by passing to a slightly abstract real space formulation in non-commutative space without choosing 
any basis for the Landau level. 
To that end we {\em define}  the composite fermion field $c(\v R, \tau)$
as a function of the non-commutative guiding center coordinate $\vec R$, and imaginary time $\tau$: 
\bea
c(\v R, \tau) & = &  \int \frac{d^2\v k}{(2\pi)^\frac{1}{2}} e^{i\vec k \cdot \vec R} c_{\v k, \tau} \\
\overline{c}(\v R, \tau) & = &  \int \frac{d^2\v k}{(2\pi)^\frac{1}{2}} e^{- i\vec k \cdot \vec R} \overline{c}_{\v k, \tau} 
\eea
As emphasized in Sec. \ref{ncft}  functions of $\v R$ should be viewed as operators that act within the space of single particle LLL states. In a basis $\{|m\rangle\}$ for these single particle  LLL states, the matrix elements of the (Grassmann-valued) function $c(\v R, \tau) $ are 
\be
\langle m|c(\v R, \tau) |n \rangle = \int \frac{d^2\v k}{(2\pi)^\frac{3}{2}} \langle m| e^{i\vec k \cdot \vec R}|n \rangle  c_{\v k, \tau}
\ee
Thus we see that these matrix elements are precisely the Grassmann fields $c_{mn}$ corresponding to the matrix-values composite fermion operators we have been working with. 

As explained in Sec. \ref{ncft} such non-commutative fields can be traded for  fields $c(\v x)$ defined in ordinary space $\v x$ so long as we modify products to star products. We will use this formulation below.   
 
 In terms of these fields the mean field action becomes 
 \be
 \label{nccflmflag}
 {\cal S}_{HF} = \int d\tau d^2 \v x ~~
 \left( \overline{c}(\v x, \tau)* \frac{d c_{\v x}}{d\tau} + \frac{1}{2m}\v \nabla \overline{c}(\v x, \tau)* \v \nabla c(\v x , \tau)    \right)
 \ee

The `right' and `left' gauge transformations of  Eqns. \ref{rightg} and \ref{leftg} act on $c(\v x, \tau)$ through 
\be
\label{ncgaugeT}
c(\v x, \tau) \rightarrow U^L(\v x, \tau)*c(\v R, \tau)* U^R(\v x, \tau) 
\ee
with $U^{L,R} = e^{i\theta_{L,R}(\v x, \tau)}$; the exponential is defined through it's power series with all products being star products. 

To obtain a gauge invariant effective action for the fluctuations we  introduce a dynamical non-commutative $U(1)$ gauge field $a_\mu(\v x, \tau)$ and a background $U(1)$ gauge field $A_\mu(\v x, \tau)$. Under the gauge transformation Eqn. \ref{ncgaugeT} these transform as 
\bea
a_\mu & \rightarrow & U_R^\dagger*  a_\mu *U_R + i  U_R^\dagger * \partial_\mu U_R   \\
A_\mu & \rightarrow & U_L *  A_\mu* U^\dagger_L + i \partial_\mu U_L *U^\dagger_L
\eea

We will assume that $a_\mu$ and $A_\mu$ are both slowly varying on the scale of the magnetic length. They thus respond to 
long wavelength gauge transformations which is what we are interested in. The important components of the fermion fields are however 
not at low momenta but rather at momenta close to the Fermi surface. 

We now  recall the covariant derivatives introduced in  Eqn. \ref{nccdrv}
\be 
D_\mu c = \partial_\mu c - ic *a_\mu - i A_\mu*c 
\ee

Under the gauge transformations of Eqn. \ref{ncgaugeT}, these derivatives transform as 
\be
D_\mu c \rightarrow U^L(\v x, \tau)*D_{\mu}c(\v x, \tau)* U^R(\v x, \tau) 
\ee

It will be useful below to also note the infinitesimal form of these gauge transformations. Under an infinitesimal right gauge transformation $U_R = 1 + i\theta_R$
we have 
\begin{eqnarray}
\label{smallRgc}
c & \rightarrow & c + i c*\theta_R \\
\label{smallRga}
a_\mu & \rightarrow & a_\mu + \partial_\mu \theta_R + i(a_\mu *\theta_R - \theta_R * a_\mu) 
\end{eqnarray}
and under an infinitesimal left gauge transformation $U_L = 1 + i\theta_L$, we have 
\begin{eqnarray}
\label{smallLgc}
c & \rightarrow & c + i \theta_L * c \\
\label{smallLgA}
A_\mu & \rightarrow & A_\mu + \partial_\mu \theta_L + i(\theta_L*A_\mu - A_\mu * \theta_L) 
\end{eqnarray}

We can now construct an effective action that correctly captures the effect of gauge fluctuations about the mean field state; we simply replace all the derivatives in Eqn.~\ref{nccflmflag} by covariant derivatives to get Eqn.~\ref{nccfllag}: 
\be
\label{ncft_cfl}
{\cal S} = \int d^2\v x d\tau~~ \overline{c} * D_0 c + i a_0 \underline \rho +  \frac{1}{2m^*} \overline{D_i c}{D_i c}  
\ee
Thus the formulation as a non-commutative field theory readily allows us to  identify and to 
formulate a theory of the important fluctuations beyond mean field. The fermions are at the non-zero density $\underline{\rho}$.
As promised this is a theory of a Fermi surface coupled to a $U(1)$ gauge field without a Chern-Simons term; 
however the theory is defined in non-commutative space.

\section{Approximation as a commutative field theory}
\label{sec:comm_approx}

The non-commutative  effective field theory is the natural result of describing the composite fermi liquid in the lowest Landau level. It is hard to directly compare it to other proposed field theories for this composite fermi liquid which are defined in commutative space (corresponding to the absence of Landau level projection).  However  the non-commutativity of space occurs at the scale of the magnetic length $l_B$, and we might suppose that for fluctuations at a wavelength much bigger than $l_B$ there is an approximate commutative effective field theory.  Interestingly precisely such an approximate mapping between non-commutative and commutative gauge theories was discovered in a well-known paper by Seiberg and Witten\cite{Seiberg_1999}. Here we extend the  Seiberg-Witten map to include fermion fields, and apply it to the field theory of the composite fermi liquid.  This will enable us to obtain an approximate commutative field theory for long wavelength fluctuations of the composite fermi liquid. 

\subsection{The Seiberg-Witten map}
\label{sec:swmap}

The Seiberg-Witten map is usually presented as an expansion of the non-commutative fields and gauge transformation parameters in powers of the non-commutativity parameter $\Theta = - l_B^2$. The coefficients in this expansion are expressed in terms of commutative fields and corresponding gauge transformation parameters. We will follow this presentation here. 

We formally seek a map from non-commutative fields and gauge transformation parameters $(a_\mu, A_\mu, c, \theta_R, \theta_L)$ to commutative fields and gauge transformation parameters denoted $(\hat{a}_\mu, \hat{A}_\mu, \psi, \hat{\theta}_R, \hat{\theta}_L)$ of the form:
\begin{eqnarray}
a_\mu &= &  a_\mu(\hat{a}_\nu) \\
A_\mu & = & A_\mu(\hat{A}_\nu) \\
c & = & c(\psi, \hat{a_\mu}, \hat{A_\mu}) \\
\theta_R & = & \theta_R(\hat{a}_\mu, \hat{\theta_R}) \\
\theta_L & = & \theta_L(\hat{A}_\mu, \hat{\theta_L}) 
\end{eqnarray}
We require that the hatted fields satisfy the standard commutative gauge transformations: 
\begin{eqnarray}
\label{smallcmmgt}
\hat{a}_\mu & \rightarrow & \hat{a}_\mu + \partial_\mu \hat{\theta}_R \\
\hat{A}_\mu & \rightarrow & \hat{A}_\mu + \partial_\mu \hat{\theta}_L \\
\psi & \rightarrow & \psi + i\psi (\hat{\theta}_L + \hat{\theta}_R)
\end{eqnarray}
It is a priori not clear that such a map will exist but we will find it explicitly to linear order in $\Theta$.  Furthermore the map will determine the non-commutative fields at a space-time point in terms of the commutative fields (and their derivatives) at the same point. Finally note that the Seiberg-Witten map relates the gauge transformation parameters in a manner that depends on the gauge field configurations.

The assumption that the map is analytic around $\Theta=0$ allows us to write it down as
\begin{align}
\label{mapexp}
    A(\hat{A})&=\hat{A}+\Delta A(\hat{A}) & a(\hat{a})&=\hat{a}+\Delta a(\hat{a}) \nonumber\\
    \theta_L(\hat{\theta}_L,\hat{A})&=\hat{\theta}_L+\Delta \theta_L(\hat{\theta}_L,\hat{A}) & \theta_R(\hat{\theta}_R,\hat{a})&=\hat{\theta}_R+\Delta\theta_R(\hat{\theta}_R,\hat{a})\\
    c(\psi,\hat{A},\hat{a})&=\psi+\Delta \psi(\psi,\hat{A},\hat{a})  \nonumber
\end{align}
Here $\Delta A$, $\Delta a$, $\Delta \theta_R$, $\Delta \theta_L$, and $\Delta c$ are all of $o(\Theta)$.

To determine the map, we start with two sets of fields in commutative space $(\hat{A},\hat{a},\psi)$ and $(\hat{A'},\hat{a'},\psi')$, which differ by a standard infinitesimal gauge transform in Eqn.~\ref{smallcmmgt}. The map, if it exists, should send them respectively to two sets of fields in non-commutative space $(A,a,c)$ and $(A',a',\psi)$, which by assumption are also connected through a gauge transform in non-commutative space as described by Eqn.~\ref{smallRgc}, \ref{smallRga} and \ref{smallLgc}, \ref{smallLgA}. Moreover, the two gauge transformation parameters for the commutative fields and their non-commutative counterparts are related by the second line of Eqn.~\ref{mapexp}. The constraints are written as
\begin{eqnarray}
\label{cnstr}
    A_\mu(\hat{A}+\partial\hat{\theta}_L)-A_\mu(\hat{A})&=&+\partial_\mu \theta_L(\hat{\theta}_L,\hat{A})
    + i\left[\theta_L(\hat{\theta}_L,\hat{A}),A_\mu(\hat{A})\right]_* \nonumber\\
    a_\mu(\hat{a}+\partial\hat{\theta}_R)-a_\mu(\hat{a})&=&+\partial_\mu \theta_R(\hat{\theta}_R,\hat{a})
    + i\left[a_\mu(\hat{a}),\theta_R(\hat{\theta}_R,\hat{a})\right]_* \\
    c\left(\psi + i\psi (\hat{\theta}_L + \hat{\theta}_R),\hat{A}+\partial\hat{\theta}_L,  \hat{a}+\partial\hat{\theta}_R\right)&-&c\left(\psi,\hat{A},\hat{a}\right) \nonumber\\
    & = & i\left(\theta_L(\hat{\theta}_L,\hat{A})*c(\psi,\hat{A},\hat{a})
    +  c(\psi,\hat{A},\hat{a}) * \theta_R(\hat{\theta}_R,\hat{a})\right) \nonumber
\end{eqnarray}
where we have used a shorthand notation $\left[A,B\right]_*=A*B-B*A$. Now, Eqn.~\ref{cnstr} can be expanded using Eqn.~\ref{mapexp}. By comparing terms up to first order in $\theta_L$, $\theta_R$ and $\Theta$, one can show the map is
\begin{align}
\label{eq:map}
     \Delta A_{\mu}(\hat{A})&=-\frac{\Theta}{2}\epsilon^{\nu \rho}\hat{A}_{\nu}(\partial_{\rho} \hat{A}_{\mu}+\hat{F}_{\rho \mu}),
    & \Delta a_{\mu}(\hat{a})&=+\frac{\Theta}{2}\epsilon^{\nu \rho}\hat{a}_{\nu}(\partial_{\rho} \hat{a}_{\mu}+\hat{f}_{\rho \mu}) \nonumber\\
     \Delta\theta_L(\hat{\theta_L},\hat{A})&=-\frac{\Theta}{2}\epsilon^{\mu \nu}\hat{A}_{\mu}\partial_{\nu}\hat{\theta}_L,
    & \Delta\theta_R(\hat{\theta}_R,\hat{a})&=+\frac{\Theta}{2}\epsilon^{\mu\nu}\hat{a}_{\mu}\partial_{\nu}\hat{\theta}_R \\
     \Delta \psi(\psi,\hat{A},\hat{a})&=+\frac{\Theta}{2}\epsilon^{\mu\nu}\left[(\hat{a}_{\mu}-\hat{A}_{\mu})\partial_{\nu}\psi-i\hat{a}_\mu\hat{A}_\nu\psi\right] \nonumber
\end{align}
where $\epsilon^{12}=-\epsilon^{21}=1$, and $\epsilon^{\mu \nu}=0$ for all other entries. $\hat{F}_{\mu \nu}=\partial_{\mu}\hat{A}_{\nu}-\partial_{\nu}\hat{A}_{\mu}$, $\hat{f}_{\mu \nu}=\partial_{\mu}\hat{a}_{\nu}-\partial_{\nu}\hat{a}_{\mu}$ are the field strengths.

\subsection{Emergence of  HLR theory}
\label{sec5A}
 Using the Seiberg-Witten map, we can rewrite the non-commutative Lagrangian in Eqn. \ref{ncft_cfl} in terms of the fields defined in commutative space to obtain a Lagrangian, formally to linear order in $\Theta$. Let us write the resulting Lagrangian as 
\be
{\cal L} = {\cal L}_0 + {\cal L}_1
\ee
where ${\cal L}_0$ is formally of order $\Theta^0$ and ${\cal L}_1$ is formally of order $\Theta$. 

We then have 
\be
{\cal L}_0 =  \bar{\psi} \partial_0 \psi - i (\hat{a}_0 + \hat{A}_0) \bar{\psi}\psi + i \hat{a}_0 \underline{\rho} 
 + \frac{1}{2m^*} \bigg|\left(\partial_i - i(\hat{a}_i + \hat{A}_i)\right) \psi\bigg|^2 
\ee
${\cal L}_1$  has contributions coming from several pieces of the non-commutative Lagrangian.  We begin with the term $ i a_0 \underline{\rho}$. Using the Seiberg-Witten map we see that the $o(\Theta)$ piece from this is 
\be
i\frac{\underline{\rho}\Theta}{2} \epsilon^{\alpha\beta} \hat{a}_\alpha (\partial_\beta \hat{a}_0 + \hat{f}_{\beta 0} )
\ee
It is readily seen to be the   Chern-Simons term (after an integration by parts) 
\be
i\frac{\underline{\rho}\Theta}{2}\epsilon^{\alpha\beta\gamma}\hat{a}_\alpha\partial_\beta \hat{a}_\gamma
\ee
Note that as $\Theta = -l_B^2$ and $\underline{\rho} = \frac{1}{2\pi l_B^2}$, the coefficient of the Chern-Simons term is precisely\footnote{ Note that we have been obtained a coefficient of $o(1)$ from a term that is formally of order $\Theta$. This is because we are at a density of composite fermions that is order $\frac{1}{|\Theta|}$. We will return to this point at the end of this section.} $-\frac{1}{4\pi}$.   

Next consider the contribution at $o(\Theta)$ from the term involving the covariant 
derivative. We split this into two parts coming from the two terms in the Seiberg-Witten map for the fermion fields 
\begin{eqnarray}
 \Delta \psi(\psi,\hat{A},\hat{a})&= & \Delta^{(1)}\psi + \Delta^{(2)}(\psi) \nonumber \\
 \Delta^{(1)}\psi & = & +\frac{\Theta}{2}\epsilon^{\mu\nu}(\hat{a}_{\mu}-\hat{A}_{\mu})\partial_{\nu}\psi \nonumber \\
 \Delta^{(2)}\psi & = &  - i\frac{\Theta}{2} \epsilon^{\nu \mu} \hat{a}_\mu\hat{A}_\nu\psi
\end{eqnarray}
Thus we write 
\be
\bar{c}*D_0 c = \bar{\psi} \hat{D_0} \psi + {\cal L}_{\tau, 1} + {\cal L}_{\tau, 2}
\ee
where the ${\cal L}_{\tau,2}$ term comes from $\Delta \psi^{(2)}$ and ${\cal L}_{\tau,1}$ represents the remaining contributions. We also define  $\hat{D}_\mu = \partial_\mu - i(\hat{a}_\mu + \hat{A}_\mu)$ as the standard covariant derivative for the commutative fields. We then have 
\be
\label{L1_0}
  \begin{split} {\cal L}_{\tau,1} = 
    -\frac{\Theta}{2} \epsilon^{\alpha\beta} &\bigg\{ \partial_0(\hat{a}_\beta - \hat{A}_\beta)\bar{\psi} \partial_\alpha \psi + \partial_\beta(\hat{a}_0 - \hat{A}_0) (\partial_\alpha \bar{\psi}~\psi) \\
   & + (\hat{a}_\beta - \hat{A}_\beta) \left(\partial_\alpha(\bar{\psi} \partial_0 \psi) - i(\hat{a}_0 + \hat{A}_0) \partial_\alpha(\bar{\psi}\psi) \right)\\
   & + \left[ i \hat{a}_\alpha (\partial_\beta \hat{a}_0 + \hat{f}_{\beta 0}) - i \hat{A}_\alpha (\partial_\beta\hat{A}_0 + \hat{F}_{\beta 0})\right]\bar{\psi}\psi \bigg\}
\end{split}
\ee

These terms can be simplified as we now explain. To that end we consider the equation of motion obtained from ${\cal L}_0$ by varying  the dynamical gauge fields. This gives 
\be
\label{current0}
    j^{\mu}=\frac{\delta {\cal L}}{\delta a_{\mu}}=0
\ee
    The spatial components yield the equation
\be
\label{currenti}
    \bar{\psi}\left(\hat{D}_i \psi \right)-\left(\overline{\hat{D}_i\psi}\right)\psi =0\\
\ee
 
It follows that 
\begin{eqnarray}
\bar{\psi}~\partial_i \psi & = & \frac{1}{2} \partial_i (\bar{\psi} \psi) + i (\hat{a}_i + \hat{A}_i ) \bar{\psi}\psi  \\
\partial_i\bar{\psi}~\psi & = & \frac{1}{2}  \partial_i (\bar{\psi} \psi) - i (\hat{a}_i + \hat{A}_i ) \bar{\psi}\psi  
\end{eqnarray}
We use this to reduce the first line of ${\cal L}_{\tau,1}$ in Eqn. \ref{L1_0} to 
\be
    -\frac{\Theta}{2} \epsilon^{\alpha\beta}\left[ \left(\partial_0(\hat{a}_\beta - \hat{A}_\beta) + \partial_\beta(\hat{a}_0 - \hat{A}_0)\right)\frac{1}{2} \partial_\alpha(\bar{\psi}\psi) \\
- i(\hat{a}_\alpha + \hat{A}_\alpha)\left(\partial_\beta (\hat{a}_0 - \hat{A}_0) - \partial_0(\hat{a}_\beta - \hat{A}_\beta)\right)\bar{\psi}\psi \right]
\ee
or more compactly as 
\be
 \frac{\Theta}{2}\epsilon^{\alpha\beta}  \left[\frac{1}{2}\partial_0\partial_\alpha( \hat{a}_\beta - \hat{A}_\beta )  \bar{\psi}\psi + i(\hat{a}_\alpha + \hat{A}_\alpha)\left(\partial_\beta (\hat{a}_0 - \hat{A}_0) - \partial_0(\hat{a}_\beta - \hat{A}_\beta)\right)\bar{\psi}\psi \right]
 \ee
 
The second line of Eqn. \ref{L1_0} can be written
\be
-\frac{\Theta}{2} \epsilon^{\alpha\beta}\left[(\hat{a}_\beta - \hat{A}_\beta)\partial_\alpha(\bar{\psi} \hat{D}_0 \psi) + i (\hat{a}_\beta - \hat{A}_\beta) \partial_\alpha (\hat{a_0} + \hat{A}_0 ) \bar{\psi}\psi \right]
\ee

We may now sum together the first, second, and third  lines of Eqn. \ref{L1_0}. Expanding out the resulting products of gauge fields, and using the antisymmetry of $\epsilon^{\alpha\beta}$, we find that the third line is exactly cancelled by contributions from the other two lines. The remaining terms lead to 
\be 
{\cal L}_{\tau, 1} = -\frac{\Theta}{2} \epsilon^{\alpha\beta} \left( -\frac{1}{2}\partial_0\partial_\alpha( \hat{a}_\beta - \hat{A}_\beta )  \bar{\psi}\psi + (\hat{a}_\beta - \hat{A}_\beta)\partial_\alpha(\bar{\psi} \hat{D}_0 \psi) + i \partial_0(\hat{a_\alpha}\hat{A}_\beta)\bar{\psi}\psi  \right) 
\ee
The last term is not gauge-invariant under combined gauge transformations of $\hat{a}$ and $\hat{A}$. However we show in Appendix \ref{app:L_ts} that it is exactly cancelled by ${\cal L}_{\tau,2}$. Thus we have 
\be
{\cal L}_{\tau,1} + {\cal L}_{\tau,2} = -\frac{\Theta}{2} \epsilon^{\alpha\beta} \left( -\frac{1}{2}\partial_0\partial_\alpha( \hat{a}_\beta - \hat{A}_\beta )  \bar{\psi}\psi + (\hat{a}_\beta - \hat{A}_\beta)\partial_\alpha(\bar{\psi} \hat{D}_0 \psi) \right)
\ee
In the corresponding action, we integrate the last term by parts (and throw away total derivative terms) to obtain
\be
{\cal L}_{\tau,1} + {\cal L}_{\tau,2} = \frac{\Theta}{2} \epsilon^{\alpha\beta} \left( \frac{1}{2}\partial_0\partial_\alpha( \hat{a}_\beta - \hat{A}_\beta )  \bar{\psi}\psi +\partial_\alpha  (\hat{a}_\beta - \hat{A}_\beta)\bar{\psi} \hat{D}_0 \psi \right)
\ee
  
  The spatial gradient term can be similarly handled. Details may be found in Appendix \ref{app:L_ts}. We show there that the leading order in $\Theta$ term is  
  \be
{\cal L}_{x} =  -\frac{\Theta}{2} \epsilon^{\alpha \beta} \partial_\alpha (\hat{a}_\beta - \hat{A}_\beta) \frac{1}{2m^*} |\hat{D}_i \psi|^2
\ee

 Combining all these contributions we  thus obtain (to linear order in $\Theta$) the effective commutative Lagrangian: 
 \begin{eqnarray} 
 {\cal L} & = & {\cal L}_{HLR} + {\cal L}_{corr} \\
 \label{HLR1}
 {\cal L}_{HLR} & = & \bar{\psi} \partial_0 \psi - i (\hat{a}_0 + \hat{A}_0) \bar{\psi}\psi + i \hat{a}_0 \underline{\rho} 
 + \frac{1}{2m^*} \bigg|\left(\partial_i - i(\hat{a}_i + \hat{A}_i)\right) \psi\bigg|^2  - i\frac{1}{4\pi}  \epsilon^{\alpha\beta\gamma}\hat{a}_\alpha\partial_\beta \hat{a}_\gamma \\
 {\cal L}_{corr} & = & \frac{\Theta}{2} \epsilon^{\alpha\beta} \left( -\frac{1}{2}\partial_\alpha(\hat{a}_\beta - \hat{A}_\beta) \partial_0 (\bar{\psi}\psi) + \partial_\alpha  (\hat{a}_\beta - \hat{A}_\beta)(\bar{\psi} \hat{D}_0 \psi  -  \frac{1}{2m^*} |\hat{D}_i \psi|^2) \right)
 \end{eqnarray}
 
 Remarkably the first term ${\cal L}_{HLR}$ this is precisely the HLR action for the composite fermi liquid, while the second term ${\cal L}_{corr}$ is a subleading correction . To make this identification, first note that  $\hat{A}$ represents an additional probe gauge field  on top of the basic magnetic field B that defines the Landau level. If we introduce a vector potential $\underline{A}_\mu = (0, \underline{A}_x, \underline{A}_y) $ such that 
 \be
 \v \nabla \times \underline{\v A} = B
 \ee
 then the total external gauge field is 
 \be
 A_{tot,\mu}  = \hat{A}_\mu + \underline{ A}_\mu 
 \ee
 We then have 
 \be
 {\cal L}_{HLR} = {\cal L}[\psi, \hat{a} - \underline{A} + A_{tot} ] + i \hat{a}_0 \underline{\rho} - i\frac{1}{4\pi}  \epsilon^{\alpha\beta\gamma}\hat{a}_\alpha\partial_\beta \hat{a}_\gamma
 \ee
 We similarly define a new dynamical  gauge field $a_{tot}$ through
 \be
 a_{tot,\mu} = \hat{a}_\mu - \underline{A}_\mu
 \ee
 We then get 
 \be
 {\cal L}_{HLR} = {\cal L}[\psi, a_{tot} + A_{tot}] + i a_{tot,0} \underline{\rho} - \frac{i}{4\pi}  \epsilon^{\alpha\beta\gamma}(a_{tot,\alpha} + \underline{A}_\alpha)\partial_\beta (a_{tot,\gamma} + \underline{A}_\gamma)
 \ee
 Expanding out the Chern-Simons term and using $A_{tot,0} = 0$, $\underline{\rho} = \frac{B}{2\pi}$ we get the Lagrangian
 \be
 {\cal L}_{HLR} = {\cal L}[\psi, a_{tot} + A_{tot}]  - \frac{i}{4\pi}  \epsilon^{\alpha\beta\gamma}a_{tot,\alpha}\partial_\beta a_{tot,\gamma}
 \ee
 which is the standard form of HLR.  However 
  the HLR Lagrangian is usually derived through the flux attachment procedure without invoking the projection to the LLL. The composite fermion mass appearing in the usual HLR action is the bare electron mass. Here we have derived the HLR action within the LLL. It appears as an approximation to the more microscopically correct non-commutative field theory Eqn. \ref{nccfllag}. The composite fermion  mass that appears in the HLR action thus obtained is determined by the interactions. 
  
 Let us now examine the terms in ${\cal L}_{corr}$. In the absence of  the probe background gauge field ($\hat{A} = 0$), the first term is a coupling between the internal electric field and the density gradient. This is small so long as we limit ourselves to long wavelength density fluctuations. The second term involves corrections, of order $\delta \rho l_B^2 \ll 1$, to terms already present in ${\cal L}_{HLR}$. Here  $\delta \rho$ is the fluctuation of the density in real space. (We used the relationship $\epsilon^{ij} \partial_i \hat{a}_j = 2\pi \delta \rho $  implied by ${\cal L}_{HLR}$). Thus this is small with the further assumption that we limit ourselves to small amplitude fluctuations of the density. 
 
 Thus the HLR Lagrangian emerges as an approximate description of the full non-commutative field theory for long wavelength, low amplitude gauge fluctuations. 
 The crucial  Chern-Simons term arises with a properly quantized coefficient $-\frac{1}{4\pi}$.  Does the presence of a mean density of order $\frac{1}{|\Theta|}$ invalidate the expansion in powers of $\Theta$?  The mean density sets the 
 Fermi momentum $K_F \sim \frac{1}{\sqrt{|\Theta|}}$. 
 Clearly we can not assume that the fermions are at long wavelength though the important gauge fluctuations 
 are at long wavelength. It is thus reassuring that  the smallness of the correction terms in ${\cal L}_{corr}$ only invoked the long wavelength, low amplitude limit for the gauge fluctuations.  This then is a justification of the use of HLR theory for many physical properties (eg, the compressibility, or transport in the presence of a smooth impurity potential) even when restricted to the LLL.  If however we are interested in universal short-distance properties, such as $2K_F$ singularities in density correlations, it may be  safer to go back to the full non-commutative field theory. 
 
 Another consequence of the emergence of the Chern-Simons term with an $o(1)$ coefficient is that we must 
 re-examine Eqn. \ref{currenti} for the current that we used to obtain the $o(\Theta)$ correction to the action. 
 The Chern-Simons term will lead to an extra `Hall current' contribution to this equation which will lead to an 
 additional correction to the HLR action. We do this in Appendix and show that this extra correction is of the form 
 \be
\label{L2}
    -\frac{\Theta}{4\pi}m^*\left((\hat{f}_{01}-\hat{F}_{01})\hat{f}_{01}+(\hat{f}_{02}-\hat{F}_{02})\hat{f}_{02}\right)
\ee
 This is an innocuous correction for long wavelength gauge fluctuations.

Can we understand why at the end of the day we only obtain a self Chern-Simons term for $\hat{a}$? In particular based on the interpretation of the composite fermion as a vortex one might have expected a mutual Chern-Simons term of the form $\frac{i}{2\pi} \hat{A} \wedge d\hat{a}$ which is not found in our derivation. To understand this we note that the non-commutative Lagrangian, apart from the term $-ia_0\underline{\rho}$ that comes from Lagrangian multiplier for the gauge constraint Eqn.~\ref{phrcnstr}, has a symmetry  $\hat{A}\leftrightarrow \hat{a}, \Theta \rightarrow -\Theta$. This symmetry of part of the Lagrangian precludes any mutual Chern-Simons term between $\hat{a}$ and $\hat{A}$.  
The term $-ia_0\underline{\rho}$ contributes only an internal Chern-Simons term $\sim a \wedge da$.

A few qualitative (and somewhat heuristic) remarks on  the results of this section may be useful.  A well known way to understand the usual HLR  construction (without the LLL restriction) for bosons at $\nu = 1$ is in terms of a traditional parton representation  where we write the microscopic boson operator $b$ as a product of two fermions: 
\be
b = \psi f
\ee
This comes with a $U(1)$ gauge constraint $\psi^\dagger \psi = f^\dagger f$. This introduces a $U(1)$ gauge field. Further the total number of $f$ (or $\psi$ particles) equals the total number of bosons. 
We assume that the $\psi$-fermions carry the global $U(1)$ charge of the boson and see the external magnetic field. The $f$-fermions then are neutral under the global $U(1)$. In a mean field description of the composite Fermi liquid, there is a mean internal gauge flux that cancels the external gauge flux. Then the $\psi$-fermions see net effective zero magnetic field and form a Fermi surface while the $f$-fermions are in an integer quantum Hall state with $\sigma_{xy} = 1$. Integrating out the $f$-fermions, we get the standard HLR action with a Chern-Simons term for the fluctuations of the internal gauge field. Now, the $c$-fermions occuring in the PHR formulation may roughly be thought of as the LLL version of the $\psi$-fermions in the standard parton construction. The constraint that the `right' density does not fluctuate may be represented formally by introducing a filled Landau level of $f$-fermions and writing 
\be
\rho^R_{nn'} = f^\dagger_n f_{n'} 
\ee
where $f_n$ destroys an $f$-fermion in the Landau orbital $n$. It is natural then the contribution of the 
background density (which technically is the origin of the Chern-Simons term) gives a Chern-Simons term. 

Finally we briefly comment on the relationship to the ideas of Ref. \onlinecite{wang2016composite} on the emergent Berry phase of composite fermions in the LLL. That paper proposed that as the LLL limit was taken the
composite fermions of the HLR theory will develop a Fermi surface Berry phase of $-2\pi$ (for bosons at $\nu = 1$).  This Berry phase will then give an anomalous Hall effect for the internal gauge field that excatly cancels the Chern-Simons term of the original HLR theory. This then was suggested to be a way to reconcile the two effective Lagrangians discussed in Appendix \ref{app:priorftcfl}. The detailed analysis presented here partially  supports this proposal but also shows its limitation. The correct effective non-commutative field theory in the LLL has no Chern-Simons term, but the `right' density operator expressed in terms of the composite fermions has a form factor $e^{-\frac{i}{2} {\v k \times \v q}}$. Considering this for  small $|\v q|$, we can think of this form factor as describing a Berry connection ${\cal \v A}(\v k)$ in momentum space:
\be
{\cal A}(\v k) = -\frac{1}{2} \hat{z} \times \v k
\ee
The corresponding Berry curvature is 
\be
{\cal B} = -1
\ee
Thus we could say that the Chern-Simons term of the HLR theory has been accomodated instead by an anomalous Hall effect that will result from the form factor associated with the composite fermion density in the LLL theory.
However the full structure that results in the LLL is the non-commutative field theory, and not the commutative effective field theory of Eqn. \ref{vcfl}. In the commutative approximation to the full non-commutative field theory, the density operator has no non-trivial form factor (and hence no Berry phase). The Seiberg-Witten map trades the theory of fermions with a gauge field coupling to densities with a non-trivial form factor to a theory of different fermions with a gauge field coupling to densities without such a form factor but with a Chern-Simons term.

\section{Doping the composite fermi liquid: the Jain states} 
Apart from their intrinsic interest composite Fermi liquids also play a crucial role as parent states of the Jain series of gapped quantum Hall states at nearby fillings. For the bosonic composite fermi liquid at $\nu = 1$, the nearby Jain states occur at a filling $\frac{p}{p+1}$ with $p$ a large integer of either sign. Topological aspects of the Jain states are described by multi-component abelian Chern-Simons gauge theories. These Topological Quantum Field Theories of course do not capture dynamical aspects of the state, for instance, the quasiparticle gaps or details of the magnetoroton mode, etc. However for large $|p|$, both topological and some dynamical properties are universally determined by properties of  the composite Fermi liquid at $\nu = 1$. Thus, armed as we are, with a LLL theory of the composite Fermi liquid we can obtain a LLL description of the large $|p|$ Jain states.  This is not straightforward directly in the original Pasquier-Haldane-Read framework: moving away from $\nu = 1$ requires using rectangular matrices $c_{mn}$ which leads to technical complications. However  the effective field theory description readily allows us to dope away from $\nu =1$. 

To that end it is simplest to just use the approximate mapping to the commutative theory described in the last section. If we initially ignore  the extra ${\cal L}_{corr}$ term, then there is no difference with the usual HLR theory. Moving away from $\nu = 1$ by changing the external magnetic field at fixed boson density, we have 
\be
\epsilon_{ij} \partial_i \hat{ A}_j = \delta B
\ee
The internal magnetic field $\hat{b} = \epsilon_{ij} \partial_i \hat{ a}_j$ has an average value 
\be
\langle \v b \rangle = 2\pi\left(  \langle \bar{\psi}\psi \rangle - \underline{\rho}\right)  ~~ = 0
\ee
As usual the {\em net} average magnetic field seen by the composite fermions is 
\be
B^* = \delta B + \langle \hat{b} \rangle ~~= \delta B
\ee
Jain states form when the composite fermions fill $p$ Landau levels which happens when $\underline{\rho} = \frac{p \delta B}{2\pi} = \frac{p(B_{tot} - B)}{2\pi}$ which gives a filling $\frac{2\pi \underline{\rho}}{B_{tot}} = \frac{p}{p+1}$.  

Next consider the ${\cal L}_{corr}$ term. The potentially important effect comes from the second term. Replacing $\hat{b} - \delta B$ by its average $-\delta B$, we find the approximate Lagrangian
\be
{\cal L} = \left( 1 + \frac{\Theta \delta B}{2}\right)\bar{\psi} \hat{D}_0\psi + i \hat{a}_0 \underline{\rho} 
 + \frac{1}{2m^*}\left(1 - \frac{\Theta \delta B}{2}\right)  |\hat{D}_i \psi|^2  - i\frac{1}{4\pi}  \epsilon^{\alpha\beta\gamma}\hat{a}_\alpha\partial_\beta \hat{a}_\gamma
 \ee
 We can now redefine the $\psi$ field (and using $|\Theta \delta B| = |\frac{\delta B}{B}| \ll 1$ to set the coefficient of the time derivative to $1$: 
 \be
 \tilde{\psi} \approx \left( 1 + \frac{\Theta \delta B}{4}\right) \psi
 \ee
 
 The Lagrangian then becomes 
 \be
 \tilde{{\cal L}} =  \bar{\tilde{\psi}} \hat{D}_0 \tilde{\psi}   + i \hat{a}_0 \underline{\rho} 
 + \frac{1}{2\tilde{m}^*}|\hat{D}_i \tilde{\psi}|^2    - i\frac{1}{4\pi}  \epsilon^{\alpha\beta\gamma}\hat{a}_\alpha\partial_\beta \hat{a}_\gamma
 \ee
 Thus the effect of ${\cal L}_{corr}$ is to change the bare mass $m^*$ to $\tilde{m}^*$ given by 
 \be
 \label{meffren}
 \tilde{m}^* = m^* \left(1 - \frac{\delta B}{B}\right) 
 \ee
 The Landau level spacing of the composite fermions $\frac{\delta B}{\tilde{m}^*}$ gives a rough estimate of  the gap of the Jain state\footnote{This will be renormalized by gauge fluctuations which lead for small $\delta B$ to a singular correction to the effective mass. So the effective mass given in Eqn. \ref{meffren} may be expected to capture the correct gap in a window of small but not too small $|\delta B|$.}. Using the mean field estimate for $m^*$ from Eqn. \ref{meff} we thus get an approximate gap for the large $|p|$ Jain states:
 \be
 \Delta \approx \frac{(|\delta B|) U_0 }{1.54} \left(1 + \frac{\delta B}{B}\right) 
 \ee
 
 In the future it should be interesting to correctly obtain the coupling of the Jain states to geometry (and calculate the shift/Hall viscosity) within this framework.

\section{Spinful bosons in LLL at total filling \texorpdfstring{$\nu_T=1$}{TEXT}}
\label{sec:spinful boson in LLL}
In this section we generalize our results to a system of two-component bosons with global $U(2)$ symmetry in a magnetic field at a total filling factor $\nu_T = 1$. The physical Hilbert space is spanned by states fully symmetric under exchange of two particles: 
\be
\ket{(m_1,\sigma_1),...,(m_n,\sigma_n)}
\ee
where $m_i$ label orbitals (in some basis) in the LLL, and $\sigma_i$ is the $SU(2)$ spin of the $i$th particle. 
There  is a total  density operator $\rho^L_{\v q}$ that satisfies the GMP algebra.  In addition there is a spin density operator  $S^{L,a}(\v q)$ ($a = 1,2,3$) that satisfies the following commutation relations 
\begin{eqnarray}
\label{gmpspin}
\left  [S^{L, a}_{\v q},\rho^{L}_{\v q'}\right] & = &  2i\sin{\left(\frac{\v q \times \v q'}{2}\right)}S^{L,a}_{\v {q+q'}} \nonumber \\
\left  [S^{L, a}_{\v q},S^{L,b}_{\v q'}\right] & = &  2i\epsilon^{abc}\cos{\left(\frac{\v q \times \v q'}{2}\right)}S^{L,c}_{\v {q+q'}}
+2i\delta^{ab}\sin{\left(\frac{\v q \times \v q'}{2}\right)}\rho^{L}_{\v {q+q'}}
\end{eqnarray} 
We will consider a Hamiltonian 
\be
{\cal H} = \frac{1}{2}\int \frac{d^2\v q}{(2\pi)^2} U(\v q) \rho^L_{\v q} \rho^L_{-\v q} 
\ee
The treatment can be readily generalized to a more general $U(2)$ symmetric  Hamiltonian that includes, for example, an interaction between the spin densities. 

This system has been studied numerically in Refs. \onlinecite{wu2015emergent,geraedts2017emergent} for a contact interaction and there is evidence for a spin-unpolarized composite fermi liquid. Below we will provide an analytic microscopic theory\footnote{It is also easy to treat  $N$-component bosons with global $U(N)$ symmetry at a total filling $\nu_T = 1$ for general $N$ but we will not do so here.}. 

\subsection{Pasquier-Haldane construction}
The Pasquier-Haldane construction introduced in Sec.~\ref{sec:parton construction} can be naturally generalized to include spin by 
introducing spinful composite fermion $c_{\sigma,mn}$ that satisfy anti-commutation relations
\be
\{c_{\sigma,mn},c^\dagger_{n'm',\sigma'}\}=\delta_{\sigma,\sigma'}\delta_{mm'}\delta_{nn'}
\ee
Many body states in the physical Hilbert space  are then represented by 
\be
\ket{(m_1,\sigma_1),...,(m_n,\sigma_n)}=\epsilon^{n_1,...,n_N}c^\dagger_{n_1;\sigma_1,m_1}...c^\dagger_{n_Nm_N,\sigma_N}\ket{0}
\ee
The antisymmetrization over internal indices $n_i$ means that physical states are singlets under the $SU(N)$ `right' transformations generated by 
$\rho^R_{nn'} - \delta_{nn'}$ where   $\rho^R_{nn'}\ket{\psi}=\delta_{nn'}\ket{\psi}$, where the right density is now
\be
    \rho^R_{nn'}=\sum_{m\sigma}c^\dagger_{nm,\sigma}c_{\sigma,mn'}
\ee
Thus we have the constraint 
\be
\rho^R_{nn'}\ket{\psi_{phys}}=\delta_{nn'}\ket{\psi_{phys}}
\ee
We can now go to momentum space using the plane wave operators $e^{i\v q \cdot \v R}$. It is readily checked that the 
$\rho^L_{\v q}, S^{L,a}_{\v q}$ satisfy the commutation algebra of Eqns.~\ref{gmpspin}. Furthermore, just as before, 
the right density operator $\rho^R$ satisfies the GMP algebra but with the opposite sign from $\rho^L$. 
The $\rho^R$ also commute with $\rho^L, S^{L,a}$. 

We note that we can define a  right spin density operator
\be
\label{rs}
    S^{R,a}_{nn'}=\frac{1}{2}\sum_{ss'm}c^\dagger_{nm,s}\sigma^a_{ss'}c_{s,mn'}
\ee
which also has vanishing matrix elements between physical states. To show this, consider a matrix element of the commutator of right spin density and right density. By virtue of the gauge constraint Eqn.~\ref{gc},
\be
    \bra{\psi_1}[S^{R,a}_{\v q},\rho^{R}_{\v q'}]\ket{\psi_2}=0 \text{ for $\forall \v q'\neq 0$}
\ee
where $\ket{\psi_1},\ket{\psi_2}$ are physical states that satisfy $\rho^{R}_{\v q}\ket{\psi_{1,2}}=0$ for $ \forall q\neq0$.  It follows therefore from the commutation algebra in Eqn. \ref{gmpspin} that
\be
    \bra{\psi_1}s^{R,i}_{\v q}\ket{\psi_2}=0 \text{ for $\forall \psi_1, \psi_2, q\neq0$}
\ee
However, unlike $\delta \rho^R$, the operator $S^{R,a}$ (for $\v q \neq 0$) does not simply annihilate physical states. Rather it takes physical states to unphysical states. We illustrate this with an explicit example in Appendix \ref{app:rsd}.    
So the right spin density is not a generator for gauge fluctuations, and the gauge structure of our spinful composite fermion construction is still $SU(N)$.

\subsection{Hartree-Fock theory}
\begin{figure}
    \centering
    \includegraphics[width=0.48 \textwidth]{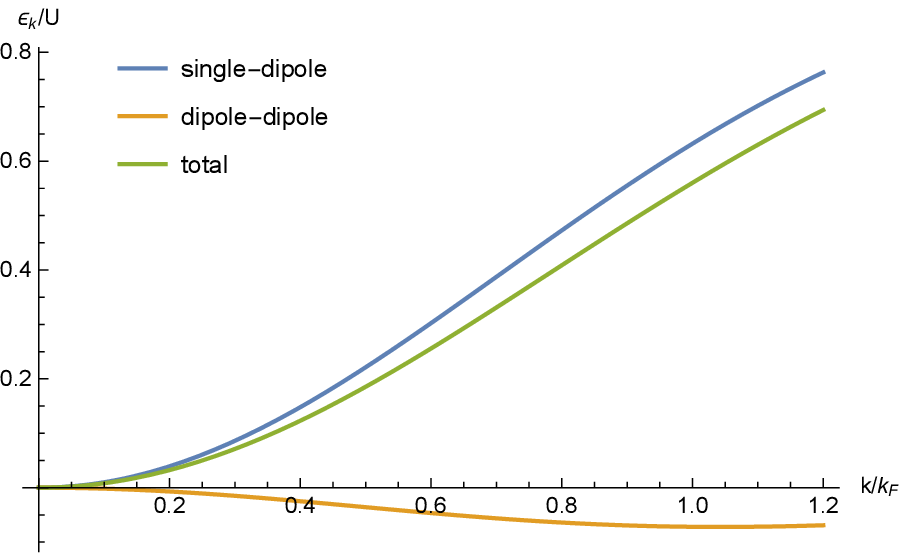}
    \includegraphics[width=0.48 \textwidth]{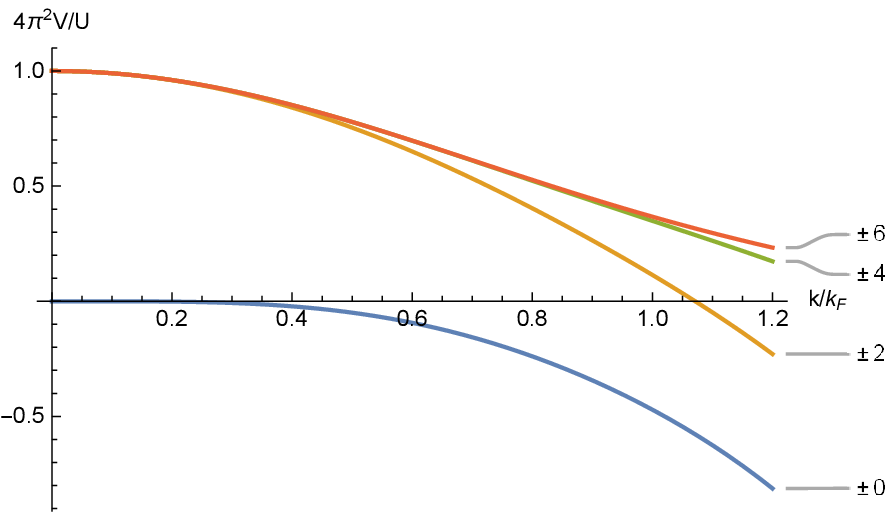}
    \caption{Mean-field dispersion and partial wave components for different angular momentum. Horizontal axis show $k/k^s_F$. 
    (a) Composite fermion dispersion at mean-field level without pairing included. The blue and orange curves denote contributions 
    from the ``single-dipole" and from the ``dipole-dipole" terms, respectively. The green curve is their sum, and hence the total dispersion. 
    (b) Partial wave components for even parity pairing channels, as in Eqn.~\ref{eq:partial wave}. The labels stand for corresponding angular momentum $l$. At the Fermi surface of spinful composite fermion, s-wave channel is attractive.}
    \label{fig:mean-field dispersion&pairing_spinful}
\end{figure}

We can now proceed completely similarly to our previous discussion. The Hamiltonian is expressed in terms of the 
$c$-fermions, and the resulting 4-fermion term can be solved within a Hartree-Fock approximation. We 
first describe a spin-unpolarized composite Fermi liquid solution (no pairing terms) with 
\be
\langle c^\dagger_{\v k s}c_{\v k s'} \rangle = n_{\v k} \delta^{(2)}(\v k - \v k') \delta_{s s'} 
\ee
with $n_{\v k} = 1$ for $\v k$ inside a circular Fermi surface of radius $k_F$, and zero otherwise. The Fermi momentum $k^s_F$ is the 
one appropriate for spinful fermions, {\em i.e}, it satisfies 
\be
\frac{2 \pi (k^s_F)^2}{(2\pi)^2} = \underline{\rho} 
\ee
We then get the dispersion for the spinful composite fermion
\be
\label{eq:HF dipersion spinful}
    \tilde{\epsilon}_{\v k}=U \left(1-e^{-\frac{k^2}{2}}\right) - 2 U e^{-\frac{k^2}{2}}\int_0^{k^s_F} dk'\: k' e^{-\frac{k'^2}{2}}\left(I_0\left(kk'\right)-1\right)
\ee
 The dispersion incorporates two terms as described in the spinless case before. The first term is the 
 intra-dipole interaction, which is unchanged compared to Eqn.~\ref{eq:HF dipersion}. The second term is an 
 inter-dipole interaction, which is different from that of Eqn.~\ref{eq:HF dipersion} due to the different 
 Fermi surface structure. We plot the mean field dispersion in Fig.~\ref{fig:mean-field dispersion&pairing_spinful}. 
 We note that the dipole-dipole term is significantly weaker than that of spinless case, since the reduced size of the  
 Fermi surfaces lead to smaller dipole moments, which provide weaker screening.

Next we include the possibility of pairing to discuss the stability of the composite Fermi liquid. 
Note that compared to the spinless problem, the spin degrees of freedom allows for both even and odd angular momentum pairing.

For spin-triplet Cooper pairing, which has odd angular momentum, the pairing interaction is 
exactly the same as Fig.~\ref{fig:mean-field dispersion&pairing}(b). In this case, the pairing in $l=\pm1$ channel is no longer attractive at the reduced Fermi surface $k^s_F=k_F/\sqrt{2}$. For spin-singlet (even angular momentum) pairing,  the pairing potential is shown in Fig.~\ref{fig:mean-field dispersion&pairing_spinful}. We find the s-wave channel is attractive. The attractive potential at the Fermi surface is around $25\%$ weaker than that of the p-wave attraction for the spinless case. Thus at the mean field level the composite Fermi liquid will be unstable to pairing, and a topologically ordered ground state will result. 
Solving the Hartree-Fock equations numerically, when only triplet pairing channel is turned on, no pairing is observed. Allowing singlet pairing, the self-consistent mean field calculation converge to the s-wave pairing state, with an energy gap $
\frac{\Delta E}{UN}=6\times 10^{-6}$, an order of magnitude smaller than that of spinless case. This is consistent with our analytical results. 

However upon including fluctuations the weaker pairing in the spinfull problem may not be able to compete against the Amperean repulsion coming from the current-current interaction. In any case we expect that the pairing is likely a weaker instability than in the spinless case. 
This is qualitatively consistent with what is seen in the numerics, where the CFL state seems to exist in the spinful model for 
currently accessible system sizes while the spinless case is in a paired state. 


\subsection{Effective field  theory} 
 Now we include fluctuations  beyond Hartree-Fock to write down a low energy effective field theory for the spinful composite fermi liquid, completely parallel to what was done in Sec.\ref{sec4}.   The Hartree-Fock composite Fermi liquid state breaks the right gauge transformations generated by $\rho^R_{v q}$ for $\v q \neq 0$: 
 \be
\label{rightgspin}
c_{\sigma,mn} \rightarrow c_{\sigma,mn'} U^R_{n'n}
\ee
where $U^R$ is an $SU(N)$ matrix. Meanwhile, we also include a left gauge transformations generated by left density $\rho^L$: 

 The important fluctuations therefore are gauge fluctuations at small $|\v q|$.  As before we will include also a background gauge field that couples to `left' $SU(N)$ rotations: 
 \be
\label{leftgspin}
c_{\sigma,mn} \rightarrow U^L_{mm'}c_{\sigma,m'n}
\ee
with $U^L$ another $SU(N)$ matrix.  In principle we could also include a background gauge field that couples  to  spin (or more precisely a $U(2)$ background gauge field that couples to both charge and spin) but we will not do so here. As before these gauge fluctuations are readily incorporated in a path integral framework in terms of the non-commutative operator-valued fields
\be
c_s(\v R, \tau) = \int \frac{d^2\v k}{(2\pi)^\frac{3}{2}} e^{i\v k \cdot \v R} c_{\v k s}(\tau)
\ee
or their corresponding ordinary fields $c_s(\v x, \tau)$ which are multiplied by the star product. 
Following the development in Sec. \ref{sec4}, we find the non-commutative effective field theory 
 
\be
{\cal S} = \int d^2\v x d\tau~~ \overline{c}_s * D_0 c_s - i a_0 \underline \rho +  \frac{1}{2m^*} |D_i c_s|^2 
\ee
where the spin index $s$ is summed over.   

Finally our discussion on the Seiberg-Witten map still applies to this spinful case, only with a modification to include spin indices of composite fermion fields. Namely, we only substitute the last line of Eqn.~\ref{eq:map} with
\be
     \Delta \psi_\sigma(\psi_\sigma,\hat{A},\hat{a})=\frac{\Theta}{2}\epsilon^{\mu\nu}\left[(\hat{a}_{\mu}-\hat{A}_{\mu})\partial_{\nu}\psi_\sigma-i\hat{a}_\mu\hat{A}_\nu\psi_\sigma\right]
\ee
We then find that the non-commutative theory is mapped to a HLR theory for the spinful composite fermi liquid (with subleading correction terms similar to Sec. \ref{sec5A}): 
 \be 
 {\cal L} = \bar{\psi}_\sigma \partial_0 \psi_\sigma - i (\hat{a}_0 + \hat{A}_0) \bar{\psi}_\sigma\psi_\sigma + i \hat{a}_0 \underline{\rho} 
 + \frac{1}{2m^*} \bigg|\left(\partial_i - i(\hat{a}_i + \hat{A}_i)\right) \psi_\sigma\bigg|^2  - i\frac{1}{4\pi} \epsilon^{\alpha\beta\gamma}\hat{a}_\alpha\partial_\beta \hat{a}_\gamma
 \ee

\section{Discussion}
The non-commutative field theory formulation of the bosonic composite fermi liquid within the LLL developed in this paper raises a number of other questions. The most important one is whether for fermions at $\nu = \frac{1}{2}$ in the LLL there is a similar formulation. Such a field theory will presumably automatically incorporate particle-hole symmetry and will reduce to the commutative field theory of the Dirac composite fermion (Ref. \onlinecite{son2015composite} or the more refined version in Ref. \onlinecite{seiberg2016duality}).  Unfortunately a direct extension of the Pasquier-Haldane-Read representation (using for instance $3$-index fermionic partons (D. Green and N. Read, unpublished; see  the thesis\cite{green2002strongly}) is complicated and has not thus far led to progress. 

Other problems that could be treated within the Pasquier-Haldane-Read formalism include mutlicomponent fermions or bosons in Landau levels at total integer filling. These have been of interest in various contexts. A further generalization is to introduce some dispersion to broaden the  Landau level into a Chern band  and study the competition between correlations and bandwidth. For bosons at $\nu = 1$ we will describe this competition and the  evolution of the ground state elsewhere (Z. Dong  and T. Senthil, to appear).  
 
\section{Acknowledgement}
We particularly thank   Hoi-Chun (Adrian) Po and  Ya-Hui Zhang for many stimulating discussions. Thanks are also due to  Nick Read and Chong Wang for sharing their insights  on many matters pertinent to this paper, and to Hong Liu for  discussions on   non-commutative field theory.  This work was supported by NSF grant DMR-1911666,
and partially through a Simons Investigator Award from
the Simons Foundation to Senthil Todadri. This work was also partly supported by the Simons Collaboration on Ultra-Quantum Matter, which is a grant from the Simons Foundation (651440, TS).  
 Part of this work was performed during a visit of TS
at the Aspen Center for Physics, which is supported by National Science Foundation grant PHY-1607611.

\appendix
\section{Field theories for the bosonic composite fermi liquid}
\label{app:priorftcfl}
In this Appendix we present the field theory suggested in Ref. \onlinecite{Read_1998} and further discussed in Refs. \onlinecite{avl1,wang2016composite}. In this theory, the composite fermion field may be given an interpretation as a (``fermionized") vortex of the physical bosons. Thus we will refer to this  as the  ``Vortex Composite Fermi Liquid" (VCFL), and will denote the corresponding composite fermion field $\psi_v$.  The Lagrangian takes the form 
\be
\label{vcfl}
{\cal L}_{vcfl} = \bar{\psi}_v (\partial_\tau - i a_0) \psi_v + \frac{1}{2m^*} |\left(\partial_i - i a_i \right) \psi_v|^2 + \frac{i}{2\pi} \epsilon_{\mu\nu\lambda} A_\mu \partial_\nu a_\lambda
- \frac{i}{4\pi} \epsilon_{\mu\nu\lambda} A_\mu \partial_\nu A_\lambda
\ee
Here $a$ is the dynamical $U(1)$ gauge field and $A_\mu$ is the background $U(1)$ gauge field. 

Though this form of the action was not explicitly written down in Ref. \onlinecite{Read_1998}, the comments in Sec II.D of that paper suggested that this effective theory might describe the microscopic results in the bulk of the paper. This effective Lagrangian should be contrasted with that for the HLR theory: 
\be
{\cal L}_{HLR} = \bar{\psi} \left(\partial_\tau - i (a_0 +A_0)\right) \psi + \frac{1}{2m^*} |\left(\partial_i - i (a_i + A_i) \right) \psi|^2 - \frac{i}{4\pi} \epsilon_{\mu\nu\lambda} a_\mu \partial_\nu a_\lambda
\ee
In the microscopic derivation of HLR $m^*$ is just the bare boson mass but if this Lagrangian emerges in a LLL theory we should regard $m^*$ as a renormalized effective mass.

Both ${\cal L}_{HLR}$ and ${\cal L}_{vcfl}$ describe (possibly distinct) composite fermi liquid phases of bosons at $\nu = 1$. In both theories all local operators are bosonic; in particular the operator with charge-$1$ under the background $U_A(1)$ gauge transformation is bosonic. The physical properties (deduced within, for instance, the Random Phase Approximation) of both theories are similar and describe metallic compressible phases. Nevertheless the two Lagrangians are different and it is not clear whether they describe the same IR fixed point or not. Furthermore it has also not been clear which, if any,  of these two arises within a microscopic LLL treatment. 

If we dispense with the LLL requirement, we can understand how to obtain either of these two theories. The HLR Lagrangian can of course be obtained by a flux attachment transformation of the original boson to composite fermion variables. The VCFL theory can be obtained as follows\cite{avl1}. First perform a standard charge-vortex duality transformation of the boson system to pass to a theory in terms of (bosonic) vortices coupled to a dynamical $U(1)$ gauge field. At boson filling $\nu = 1$, the vortices are at finite density and themselves see the boson density as an effective magnetic field; the vortices are then at a filling $\nu_{vortex} = -1$. If we now do a flux attachment transformation to fermionize these vortices, we arrive at Eqn. \ref{vcfl} (up to corrections involving higher derivative terms). 

A different possible relationship between the HLR theory and Eqn. \ref{vcfl} was described in Ref. \onlinecite{wang2016composite}. These authors proposed that in the LLL limit the HLR composite fermions acquire a Fermi surface Berry phase $-2\pi$. Upon restricting to the vicinity of the Fermi surface  we should include  an anomalous Hall effect contribution to the dynamics of the combined gauge field $a+A$. This then precisely yields the vortex composite fermi liquid Lagrangian restricted to the modes near the Fermi surface.

\section{Details of the Seiberg-Witten map}
\label{app:L_ts}
Here we provide some detail that was left out in the main text on the approximate mapping of the non-commutative effective field theory to the commutative one. We will only discuss the spinless case. 

The correction to covariant time derivative term is
\be
\label{eq:L_1t}
\begin{split}
    {\cal L}^{\tau}_1
    & = \bar{c}D_0c-\bar{\psi}\hat{D}_0\psi\\
    & = \bar{\psi}\hat{D}_0\Delta\psi+\Delta\bar{\psi}\hat{D}_0\psi - i(\Delta a_0+\Delta A_0)\bar{\psi}\psi\\
    & - i \bar{\psi} \left[(\psi*\hat{a}_0-\hat{a}_0\psi)+(\hat{A}_0*\psi-\psi \hat{A}_0)\right]
\end{split}
\ee
where $\hat{D_\mu}\psi=(\partial_\mu-i\hat{a}_\mu-i\hat{A}_\mu)\psi$.
First two terms in Eqn.~\ref{eq:L_1t} give
\be
\label{eq:L_1t-a1}
\begin{split}
    {\cal L}^\tau_{1a}=\frac{\Theta}{2}\epsilon^{\alpha\beta}
    \bigg\{
    &(\hat{a}_\alpha-\hat{A}_\alpha)
    \left[\bar{\psi}\hat{D}_0(\partial_\beta\psi)+(\partial_\beta\bar{\psi})\hat{D}_0\psi\right]
    -(i\hat{a}_\alpha\hat{A}_\beta\bar{\psi}\hat{D}_0\psi+h.c.)\\
    &\partial_0(\hat{a}_\alpha-\hat{A}_\alpha)\bar{\psi}\partial_\beta\psi
    -i\partial_0(\hat{a}_\alpha\hat{A}_\beta)\bar{\psi}\psi
    \bigg\}
\end{split}
\ee
where the last two terms come from $\partial_0$ acting on $\Delta \psi$ in the first term of Eqn.~\ref{eq:L_1t}. We separate out the gauge invariant part by organizing the term (\ref{eq:L_1t-a1}) into
\be
\label{eq:L_1t-a2}
\begin{split}
    {\cal L}^\tau_{1a}=\frac{\Theta}{2}\epsilon^{\alpha\beta}
    \bigg\{
    &(\hat{a}_\alpha-\hat{A}_\alpha)\partial_\beta(\bar{\psi}\hat{D}_0\psi)
    +i(\hat{a}_\alpha-\hat{A}_\alpha)\partial_\beta(\hat{a}_0+\hat{A}_0)(\bar{\psi}\psi)\\
    &\partial_0(\hat{a}_\alpha-\hat{A}_\alpha)(\bar{\psi}\partial_\beta\psi)
    -i\partial_0(\hat{a}_\alpha\hat{A}_\beta)\bar{\psi}\psi
    \bigg\}
\end{split}
\ee
The first term is readily seen to be gauge invariant after integration by parts.
The third term in Eqn.~\ref{eq:L_1t} becomes
\be
\label{eq:L_1t-b}
{\cal L}^\tau_{1b}=\frac{\Theta}{2}\epsilon^{\alpha\beta}\left[
-i\hat{a}_\alpha(\partial_\beta \hat{a}_0+\hat{f}_{\beta0})
+i\hat{A}_\alpha(\partial_\beta \hat{A}_0+\hat{F}_{\beta0})
\right]\bar{\psi}\psi
\ee
The last term in Eqn.~\ref{eq:L_1t} is
\be
\label{eq:L_1t-c}
{\cal L}^\tau_{1c}=\frac{\Theta}{2}\epsilon^{\alpha\beta}
(\bar{\psi}\partial_\alpha\psi) \partial_\beta(\hat{a}_0-\hat{A_0})
\ee
Now we sum up Eqn.~\ref{eq:L_1t-a2}-\ref{eq:L_1t-c} and get
\be
\label{eq:L_1t-sum}
\begin{split}
    {\cal L}^\tau_{1}=\frac{\Theta}{2}\epsilon^{\alpha\beta}
    \bigg\{&+(\hat{a}_\alpha-\hat{A}_\alpha)\partial_\beta(\bar{\psi}\hat{D}_0\psi)\\
    &+i(\hat{a}_\alpha-\hat{A}_\alpha)\partial_\beta(\hat{a}_0+\hat{A}_0)\bar{\psi}\psi\\
    &+\partial_0(\hat{a}_\alpha-\hat{A}_\alpha)(\bar{\psi}\partial_\beta\psi)-(\bar{\psi}\partial_\beta\psi) \partial_\alpha(\hat{a}_0-\hat{A_0})\\
    &+\left[-i\hat{a}_\alpha(\partial_\beta \hat{a}_0+\hat{f}_{\beta 0})
    +i\hat{A}_\alpha(\partial_\beta \hat{A}_0+\hat{F}_{\beta 0})\right]\bar{\psi}\psi \bigg\}
\end{split}
\ee
Again we separate gauge invariant terms in the third line of Eqn.\ref{eq:L_1t-sum}
\be
\begin{split}
    {\cal L}^\tau_{1}=\frac{\Theta}{2}\epsilon^{\alpha\beta}
    \bigg\{
    &+(\hat{a}_\alpha-\hat{A}_\alpha)\partial_\beta(\bar{\psi}\hat{D}_0\psi)\\
    &+i(\hat{a}_\alpha-\hat{A}_\alpha)\partial_\beta(\hat{a}_0+\hat{A}_0)(\bar{\psi}\psi)
    -i\left[
    \partial_\alpha(\hat{a}_0-\hat{A_0})
    -\partial_0(\hat{a}_\alpha-\hat{A}_\alpha)
    \right]
    (\hat{a}_\beta+\hat{A}_\beta)\bar{\psi}\psi\\
    &-\left[
    \partial_\alpha(\hat{a}_0-\hat{A_0})
    -\partial_0(\hat{a}_\alpha-\hat{A}_\alpha)
    \right]
    (\bar{\psi}\hat{D}_\beta\psi)\\
    &+\left[
    -i\hat{a}_\alpha(\partial_\beta \hat{a}_0+\hat{f}_{\beta 0})
    +i\hat{A}_\alpha(\partial_\beta \hat{A}_0+\hat{F}_{\beta 0})
    -i\partial_0(\hat{a}_\alpha\hat{A}_\beta)
    \right]
    \bar{\psi}\psi
    \bigg\}
\end{split}
\ee
Integrating by parts for the first two terms, we manage to get an almost symmetric form
\be
\label{eq:L_1t-sum2}
\begin{split}
    {\cal L}^\tau_{1}=\frac{\Theta}{2}&\epsilon^{\alpha\beta}
    \bigg\{
    -\partial_\beta(\hat{a}_\alpha-\hat{A}_\alpha)(\bar{\psi}\hat{D}_0\psi)
    -\partial_\alpha(\hat{a}_0-\hat{A_0})(\bar{\psi}\hat{D}_\beta\psi)
    -\partial_0(\hat{a}_\beta-\hat{A}_\beta)(\bar{\psi}\hat{D}_\alpha\psi)\\
    &+i
    \left[
    -\partial_\beta(\hat{a}_\alpha-\hat{A}_\alpha)(\hat{a}_0+\hat{A}_0)
    -\partial_\alpha(\hat{a}_0-\hat{A_0})(\hat{a}_\beta+\hat{A}_\beta)
    +\partial_0(\hat{a}_\alpha-\hat{A}_\alpha)(\hat{a}_\beta+\hat{A}_\beta)
    \right]\bar{\psi}\psi\\
    &+i\left[
    -\hat{a}_\alpha(\partial_\beta \hat{a}_0+\hat{f}_{\beta 0})
    +\hat{A}_\alpha(\partial_\beta \hat{A}_0+\hat{F}_{\beta 0})
    +\partial_\beta\left((\hat{a}_\alpha-\hat{A}_\alpha)(\hat{a}_0+\hat{A}_0)\right)
    -\partial_0(\hat{a}_\alpha\hat{A}_\beta)
    \right]\bar{\psi}\psi
    \bigg\}
\end{split}
\ee
The first line of Eqn.~\ref{eq:L_1t-sum2} is gauge invariant. The second line becomes
\be
\label{eq:L_1t-sum2-1}
\begin{split}
    &\frac{i\Theta}{2}\epsilon^{\mu\nu\rho}
    (\hat{a}_\mu+\hat{A}_\mu)\partial_\nu(\hat{a}_\rho-\hat{A}_\rho)\bar{\psi}\psi\\
    =&\frac{i\Theta}{2}\epsilon^{\mu\nu\rho}
    (\hat{a}_\mu\partial_\nu\hat{a}_\rho-\hat{A}_\mu\partial_\nu\hat{A}_\rho-\hat{a}_{\mu}\partial_\nu\hat{A}_\rho+\hat{A}_{\mu}\partial_\nu\hat{a}_\rho)\bar{\psi}\psi\\
    =&\frac{i\Theta}{2}\epsilon^{\mu\nu\rho}
    \left[\hat{a}_\mu\partial_\nu\hat{a}_\rho-\hat{A}_\mu\partial_\nu\hat{A}_\rho+\partial_\nu(\hat{a}_\rho\hat{A}_{\mu})\right]\bar{\psi}\psi
\end{split}
\ee
The third line of Eqn.~\ref{eq:L_1t-sum2} becomes
\be
\label{eq:L_1t-sum2-2}
\begin{split}
    &\frac{i\Theta}{2}\epsilon^{\mu\nu\rho}
    (-\hat{a}_\mu\partial_\nu\hat{a}_\rho+\hat{A}_\mu\partial_\nu\hat{A}_\rho)\bar{\psi}\psi\\
    + &\frac{i\Theta}{2}\epsilon^{\alpha\beta}
    \left[-\partial_\beta(\hat{a}_\alpha\hat{a}_0)+\partial_\beta(\hat{A}_\alpha\hat{A}_0)+\partial_\beta\left((\hat{a}_\alpha-\hat{A}_\alpha)(\hat{a}_0+\hat{A}_0)\right)
    -\partial_0(\hat{a}_\alpha\hat{A}_\beta)
    \right]\bar{\psi}\psi\\
    = &\frac{i\Theta}{2}\epsilon^{\mu\nu\rho}
    (-\hat{a}_\mu\partial_\nu\hat{a}_\rho+\hat{A}_\mu\partial_\nu\hat{A}_\rho) \bar{\psi}\psi
    + \frac{i\Theta}{2}\epsilon^{\alpha\beta}
    \left[
    \partial_\beta(\hat{a}_\alpha\hat{A}_0-\hat{A}_\alpha\hat{a}_0)
    -\partial_0(\hat{a}_\alpha\hat{A}_\beta)
    \right]
    \bar{\psi}\psi\\
    = &\frac{i\Theta}{2}\epsilon^{\mu\nu\rho}
    (-\hat{a}_\mu\partial_\nu\hat{a}_\rho+\hat{A}_\mu\partial_\nu\hat{A}_\rho) \bar{\psi}\psi
    - \frac{i\Theta}{2}\epsilon^{\mu\nu\rho} \partial_\mu(\hat{a}_\nu\hat{A}_\rho)
    \bar{\psi}\psi
\end{split}
\ee
Eqn.~\ref{eq:L_1t-sum2-1} and \ref{eq:L_1t-sum2-2} cancel exactly. The remaining correction to the covariant time derivative is
\be
    {\cal L}^\tau_{1}=\frac{\Theta}{2}\epsilon^{\alpha\beta} \partial_\alpha(\hat{a}_\beta-\hat{A}_\beta)(\bar{\psi}\hat{D}_0\psi)+\frac{\Theta}{4}\partial_0(\hat{f}_{12}-\hat{F}_{12})\bar{\psi}\psi
\ee

Next we turn to the part of the action in Eqn.~\ref{ncft_cfl} involving the spatial covariant derivatives. To  first order in the non-commutativity parameter $\Theta$, we get
\be
\begin{split}
\label{L1_s}
    {\cal L}^s_1
    & = \frac{1}{2m^*}\left(|D_\alpha c|^2-|\hat{D}_\alpha \psi|^2\right)\\
    & \sim \frac{1}{2m^*}(\hat{D}_\alpha \psi)^*
    \left\{
    \hat{D}_\alpha \Delta\psi
    -i(\Delta a_\alpha+\Delta A_\alpha)\psi
    -\frac{i}{2}[\psi,\hat{a}_\alpha-\hat{A}_\alpha]_*
    \right\} + h.c.
\end{split}
\ee
Using SW map in Eqn.~\ref{eq:map}, the first term in Eqn.~\ref{L1_s} becomes
\be
\begin{split}
    {\cal L}^s_{1a}
    & = \frac{1}{2m^*}(\hat{D}_\alpha \psi)^*
    \hat{D}_\alpha \Delta\psi + h.c.\\
    & = \frac{1}{2m^*}(\hat{D}_\alpha \psi)^*
    (-\frac{\Theta}{2}\epsilon^{\beta\gamma}) \hat{D}_\alpha \left[-(\hat{a}_\beta-\hat{A}_\beta)\partial_\gamma\psi +i\hat{a}_\beta\hat{A}_\gamma\psi \right] + h.c.\\
    & = \frac{1}{2m^*}(\hat{D}_\alpha \psi)^*
    (-\frac{\Theta}{2}\epsilon^{\beta\gamma}) \bigg\{
    -(\hat{a}_\beta-\hat{A}_\beta)\partial_\gamma (\hat{D}_\alpha\psi)
    -\partial_\alpha(\hat{a}_\beta-\hat{A}_\beta)\partial_\gamma \psi
    - i(\hat{a}_\beta-\hat{A}_\beta) \partial_\gamma (\hat{a}_\alpha+\hat{A}_\alpha) \psi\\
    & +i\hat{a}_\beta\hat{A}_\gamma \hat{D}_\alpha\psi+ i\partial_\alpha(\hat{a}_\beta\hat{A}_\gamma)\psi
    \bigg\} + h.c.
\end{split}
\label{L1_sa}
\ee
The second term in Eqn.~\ref{L1_s} is
\be
\label{L1_sb}
    {\cal L}^s_{1b}
     =-\frac{1}{2m^*}(\hat{D}_\alpha \psi)^*
    i(\Delta a_\alpha+\Delta A_\alpha)\psi + h.c.
\ee
where $\Delta \hat{A}_\alpha, \Delta \hat{a}_\alpha$ are Hermitian, which will become important later. The last term in Eqn.~\ref{L1_s} gives
\be
\label{L1_sc}
\begin{split}
    {\cal L}^s_{1c}
    & =-\frac{1}{2m^*}(\hat{D}_\alpha \psi)^*
    \frac{i}{2}[\psi,\hat{a}_\alpha-\hat{A}_\alpha]_* + h.c.\\
    & = \frac{1}{2m^*}(\hat{D}_\alpha \psi)^*
    \frac{\Theta}{2}\epsilon^{\beta\gamma} \partial_\gamma (\hat{a}_\alpha-\hat{A}_\alpha) \partial_\beta\psi + h.c.
\end{split}
\ee

Eqn.~\ref{currenti_} guarantees that $(\hat{D}_\alpha \psi)^*\psi=\frac{1}{2}\partial_\alpha (\bar{\psi}\psi)$ is real. So ${\cal L}^s_{1b}$ and the third and fourth terms of Eqn.~\ref{L1_sa} are purely imaginary and get canceled by their hermitian conjugate. 
Now summing up Eqn.~\ref{L1_sa}-\ref{L1_sc}, we are left with
\be
\label{L1_s2}
\begin{split}
    {\cal L}^s_{1}
     = \frac{1}{2m^*}(\hat{D}_\alpha \psi)^*
    \frac{\Theta}{2}\epsilon^{\beta\gamma} \bigg\{
    &(\hat{a}_\beta-\hat{A}_\beta)\partial_\gamma (\hat{D}_\alpha\psi)
    + \partial_\alpha(\hat{a}_\beta-\hat{A}_\beta)\partial_\gamma \psi \\
    &+ \partial_\gamma (\hat{a}_\alpha-\hat{A}_\alpha) \partial_\beta\psi
    - i\partial_\alpha(\hat{a}_\beta\hat{A}_\gamma)\psi
    \bigg\} + h.c.
\end{split}
\ee
Thanks to Eqn.~\ref{currenti}, the last term of Eqn.~\ref{L1_s2} gets canceled by its hermitian conjugate. Upon integration by part, the first term of Eqn.~\ref{L1_s2} (+$h.c.$) becomes
\be
\begin{split}
    {\cal L}^s_{1a'}
    & = \frac{1}{2m^*}
    \frac{\Theta}{2}\epsilon^{\beta\gamma} (\hat{a}_\beta-\hat{A}_\beta)\partial_\gamma \left[(\hat{D}_\alpha \psi)^*(\hat{D}_\alpha\psi)\right] \\
    & = \frac{1}{2m^*}
    \frac{\Theta}{2}\epsilon^{\beta\gamma}\partial_\beta(\hat{a}_\gamma-\hat{A}_\gamma) |\hat{D}_\alpha \psi|^2 \\
    & = \frac{1}{2m^*}
    \frac{\Theta}{2}(\hat{f}_{12}-\hat{F}_{12}) |\hat{D}_\alpha \psi|^2
\end{split}
\ee
and the other term is
\be
\begin{split}
    {\cal L}^s_{1b'}
    & = -\frac{1}{2m^*}(\hat{D}_\alpha \psi)^*
    (\frac{\Theta}{2}\epsilon^{\beta\gamma}) 
    \left[ \partial_\alpha(\hat{a}_\gamma-\hat{A}_\gamma) - \partial_\gamma (\hat{a}_\alpha-\hat{A}_\alpha)\right] \partial_\beta\psi
     + h.c.\\
     & = -\frac{1}{2m^*}(\hat{D}_\alpha \psi)^*
    (\frac{\Theta}{2}\epsilon^{\beta\gamma}) 
    \left[ \partial_\alpha(\hat{a}_\gamma-\hat{A}_\gamma) - \partial_\gamma (\hat{a}_\alpha-\hat{A}_\alpha)\right] \hat{D}_\beta\psi
     + h.c.\\
    & = -\frac{1}{2m^*}
    \frac{\Theta}{2}\epsilon^{\beta\gamma} \left( \hat{f}_{\alpha\gamma}- \hat{F}_{\alpha\gamma} \right) (\hat{D}_\alpha \psi)^* \hat{D}_\beta\psi
     + h.c.\\
\end{split}
\ee
where in the second line we have added a vanishing term $\sim (\hat{D}_\alpha \psi)^*i(\hat{a}+\hat{A})\psi+h.c.$ to get the covariant derivative. It is easy to check that
\be
    \epsilon_{\beta\gamma}\left( \hat{f}_{\alpha\gamma}- \hat{F}_{\alpha\gamma} \right)=\left(\hat{f}_{12}-\hat{F}_{12}\right) \delta_{\alpha\beta}
\ee
Consequently
\be
    {\cal L}^s_{1} = -\frac{1}{2m^*}
    \frac{\Theta}{2}(\hat{f}_{12}-\hat{F}_{12}) |\hat{D}_\alpha \psi|^2
\ee

\section{Corrections from the Hall current}
\label{hallc}
In Eqn.\ref{currenti} assumed that the current is vanishing. However, strictly speaking,  as we discussed in the main text, 
we should include an additional Hall current coming from the Chern-Simons term in the HLR action Eqn.\ref{HLR1}. 
To be precise, the current is
\be
    J^\alpha=\frac{\delta {\cal L}}{\delta \hat{a}_\alpha}=\frac{i}{2m^*}\left(\bar{\psi}D_\alpha\psi-\overline{(D_\alpha\psi)}\psi\right)=-i\Theta\underline{\rho}\epsilon^{\alpha\mu\nu}\partial_\mu a_\nu
\ee
which is the Hall response to the internal gauge field. Therefore,
\begin{eqnarray}
\label{currenti_}
\bar{\psi}~\hat{D}_\alpha \psi & = & \frac{1}{2} \partial_\alpha (\bar{\psi} \psi) - m^*\Theta\underline{\rho}\epsilon^{\alpha\mu\nu}\partial_\mu a_\nu  \\
\overline{(\hat{D}_\alpha\psi)}~\psi & = & \frac{1}{2}  \partial_\alpha (\bar{\psi} \psi) + m^*\Theta\underline{\rho}\epsilon^{\alpha\gamma\delta}\partial_\gamma a_\delta
\end{eqnarray}
As a consequence, the correction to covariant time derivative term now becomes
\be
\begin{split}
    {\cal L}^\tau_{1}&=\frac{\Theta}{2}\epsilon^{\mu\nu\rho} \partial_\mu(\hat{a}_\nu-\hat{A}_\nu)(\bar{\psi}\hat{D}_\rho\psi)\\
    &=\frac{\Theta}{2}\epsilon^{\alpha\beta0} \partial_\alpha(\hat{a}_\beta-\hat{A}_\beta)(\bar{\psi}\hat{D}_0\psi)
    +\frac{\Theta}{2}\epsilon^{\mu\nu\alpha}\partial_\mu(\hat{a}_\nu-\hat{A}_\nu)
    \left(\frac{1}{2}\partial_\alpha(\bar{\psi}\psi)-m^*\Theta\underline{\rho}\epsilon^{\alpha\rho\sigma}\partial_\rho a_\sigma\right)\\
    &=\frac{\Theta}{2}\epsilon^{\alpha\beta} \partial_\alpha(\hat{a}_\beta-\hat{A}_\beta)(\bar{\psi}\hat{D}_0\psi)
    +\frac{\Theta}{4}\epsilon^{\alpha\beta}\partial_0\partial_\alpha(\hat{a}_\beta-\hat{A}_\beta)
    -\frac{\Theta^2}{2}m^*\underline{\rho}~\epsilon^{\alpha\mu\nu}\partial_\mu(\hat{a}_\nu-\hat{A}_\nu)
    \epsilon^{\alpha\rho\sigma}\partial_\rho a_\sigma
\end{split}
\ee
The additional term is as stated in eqn. \ref{L2}: 
\be
    -\frac{\Theta^2}{2}m^*\underline{\rho}((\hat{f}_{01}-\hat{F}_{01})\hat{f}_{01}+(\hat{f}_{02}-\hat{F}_{02})\hat{f}_{02})
\ee
which formally is of second order in $\Theta$. However as $\underline{\rho} = \frac{1}{2\pi|\Theta|}$, it really is of 
order $\Theta$. For the spatial covariant derivative terms, no correction shows up at this order 
since in Appendix.\ref{app:L_ts} we have only used the fact that $\bar{\psi}D_\alpha\psi$ is real, which is still the case. 
Note that Eqn.~\ref{L2} is not the full correction for the action to $o(\Theta^2)$ since we have only kept $o(\Theta)$ terms in Seiberg-Witten map as well as later in the expansion of the action. It is however the only term of order $\Theta^2 \underline{\rho}$.

\section{ ``Right"  spin density}
\label{app:rsd}
In this section we show that the right spin density defined in Eqn.~\ref{rs} does not annihilate all physical states. We write the generalized density operator as
\be
    \rho^{R,\alpha}_{nn'}=\sum_{m,ss'}c^\dagger_{n,ms}\sigma^\alpha_{s,s'}c_{s'm,n'}
\ee
where $\sigma^\alpha=(1,\Sigma^x/2,\Sigma^y/2,\Sigma^z/2)$, and $\Sigma^i$ are the Pauli matrices. Then $\alpha=0$ corresponds to the right density operator and $\alpha=1,2,3$ correspond to the right spin density operators defined in the main text.
The physical Hilbert space is spanned by states
\be
    \ket{\psi_{phys,m_i,s_i}}=\epsilon^{n_1n_2...n_N}c^\dagger_{n_1,m_1s_1}c^\dagger_{n_2,m_2s_2}...c^\dagger_{n_N,m_Ns_N}\ket{0}
\ee
Applying right density on a physical state, we get
\be
\begin{split}
    \rho^{R,\alpha}_{nn'}\ket{\psi_{phys,m_i,s_i}}
    &=\epsilon^{n_1n_2...n_N}c^\dagger_{n,ms}\sigma^\alpha_{s,s'}c_{s'm,n'}c^\dagger_{n_1,m_1s_1}c^\dagger_{n_2,m_2s_2}...c^\dagger_{n_N,m_Ns_N}\ket{0}\\
    &=\sum_j(-1)^{(j-1)}\epsilon^{n_1n_2...n_N}\delta_{n'n_j}\delta_{s's_j}\delta_{m,m_j}c^\dagger_{n,ms}\sigma^\alpha_{s,s'}\prod_{i\neq j}c^\dagger_{n_i,m_is_i}\ket{0}\\
    &=\sum_j(-1)^{(j-1)}\epsilon^{n_1n_2...n_N}\delta_{n'n_j}\sigma^\alpha_{s,s_j}c^\dagger_{n,m_js}\prod_{i\neq j}c^\dagger_{n_i,m_is_i}\ket{0}\\
\end{split}
\ee
where repeated indices are summed over.

It is sufficient to illustrate our point by considering a finite system and explicitly showing that the right spin density operator takes a physical state to an nonphysical state. To that end, consider $N = 2$, {\em i.e}  a system with just 2 single particle orbitals.   The ``many body"  Hilbert space is spanned by states with 2 $c$-fermions filling 8 basis states.  Consider the state 
\be
    \ket{\psi_{phys,1\uparrow2\downarrow}}=b^\dagger_{1\uparrow}b^\dagger_{2\downarrow}\ket{0}=(c^\dagger_{1,1\uparrow}c^\dagger_{2,2\downarrow}-c^\dagger_{2,1\uparrow}c^\dagger_{1,2\downarrow})\ket{0}
\ee
where $\ket{0}$ is the vacuum state of composite fermion.  This state is in the physical Hilbert space since the internal index is anti-symmetrized. Applying the right spin operator $S^z_{12}$, one gets
\be
    S^z_{12}\ket{\psi_{phys,1\uparrow2\downarrow}}=-2c^\dagger_{11\uparrow}c^\dagger_{12\downarrow}\ket{0}
\ee
which is a nonphysical state that does not get annihilated by right density $\rho^R_{nn'}$.


\providecommand{\noopsort}[1]{}\providecommand{\singleletter}[1]{#1}%

\bibliography{reference}

\begin{thebibliography}{47}
\expandafter\ifx\csname natexlab\endcsname\relax\def\natexlab#1{#1}\fi
\expandafter\ifx\csname bibnamefont\endcsname\relax
  \def\bibnamefont#1{#1}\fi
\expandafter\ifx\csname bibfnamefont\endcsname\relax
  \def\bibfnamefont#1{#1}\fi
\expandafter\ifx\csname citenamefont\endcsname\relax
  \def\citenamefont#1{#1}\fi
\expandafter\ifx\csname url\endcsname\relax
  \def\url#1{\texttt{#1}}\fi
\expandafter\ifx\csname urlprefix\endcsname\relax\def\urlprefix{URL }\fi
\providecommand{\bibinfo}[2]{#2}
\providecommand{\eprint}[2][]{\url{#2}}

\bibitem[{\citenamefont{Halperin}(2007)}]{halperincflrev}
\bibinfo{author}{\bibfnamefont{B.~I.} \bibnamefont{Halperin}},
  \emph{\bibinfo{title}{Fermion Chern-Simons Theory and the Unquantized Quantum
  Hall Effect}} (\bibinfo{publisher}{John Wiley and Sons, Ltd},
  \bibinfo{year}{2007}), chap.~\bibinfo{chapter}{6}, pp.
  \bibinfo{pages}{225--263}, ISBN \bibinfo{isbn}{9783527617258},
  \eprint{https://onlinelibrary.wiley.com/doi/pdf/10.1002/9783527617258.ch6},
  \urlprefix\url{https://onlinelibrary.wiley.com/doi/abs/10.1002/9783527617258.ch6}.

\bibitem[{\citenamefont{Willett}(1997)}]{willett97}
\bibinfo{author}{\bibfnamefont{R.~L.} \bibnamefont{Willett}},
  \bibinfo{journal}{Advances in Physics} \textbf{\bibinfo{volume}{46}},
  \bibinfo{pages}{447} (\bibinfo{year}{1997}),
  \eprint{http://dx.doi.org/10.1080/00018739700101528},
  \urlprefix\url{http://dx.doi.org/10.1080/00018739700101528}.

\bibitem[{\citenamefont{Jain}(2007)}]{jainbook}
\bibinfo{author}{\bibfnamefont{J.~K.} \bibnamefont{Jain}},
  \emph{\bibinfo{title}{Composite fermions.}} (\bibinfo{publisher}{Cambridge :
  Cambridge University Press, 2007.}, \bibinfo{year}{2007}), ISBN
  \bibinfo{isbn}{9780521862325},
  \urlprefix\url{http://libproxy.mit.edu/login?url=http://search.ebscohost.com/login.aspx?direct=true&db=cat00916a&AN=mit.001437174&site=eds-live}.

\bibitem[{\citenamefont{Halperin et~al.}(1993)\citenamefont{Halperin, Lee, and
  Read}}]{HLR}
\bibinfo{author}{\bibfnamefont{B.~I.} \bibnamefont{Halperin}},
  \bibinfo{author}{\bibfnamefont{P.~A.} \bibnamefont{Lee}}, \bibnamefont{and}
  \bibinfo{author}{\bibfnamefont{N.}~\bibnamefont{Read}},
  \bibinfo{journal}{Phys. Rev. B} \textbf{\bibinfo{volume}{47}},
  \bibinfo{pages}{7312} (\bibinfo{year}{1993}),
  \urlprefix\url{http://link.aps.org/doi/10.1103/PhysRevB.47.7312}.

\bibitem[{\citenamefont{Jain}(1989)}]{jaincf}
\bibinfo{author}{\bibfnamefont{J.~K.} \bibnamefont{Jain}},
  \bibinfo{journal}{Phys. Rev. Lett.} \textbf{\bibinfo{volume}{63}},
  \bibinfo{pages}{199} (\bibinfo{year}{1989}),
  \urlprefix\url{http://link.aps.org/doi/10.1103/PhysRevLett.63.199}.

\bibitem[{\citenamefont{Lopez and Fradkin}(1991)}]{lopez1991fractional}
\bibinfo{author}{\bibfnamefont{A.}~\bibnamefont{Lopez}} \bibnamefont{and}
  \bibinfo{author}{\bibfnamefont{E.}~\bibnamefont{Fradkin}},
  \bibinfo{journal}{Physical Review B} \textbf{\bibinfo{volume}{44}},
  \bibinfo{pages}{5246} (\bibinfo{year}{1991}).

\bibitem[{\citenamefont{Murthy and Shankar}(2003)}]{gmrsrmp03}
\bibinfo{author}{\bibfnamefont{G.}~\bibnamefont{Murthy}} \bibnamefont{and}
  \bibinfo{author}{\bibfnamefont{R.}~\bibnamefont{Shankar}},
  \bibinfo{journal}{Rev. Mod. Phys.} \textbf{\bibinfo{volume}{75}},
  \bibinfo{pages}{1101} (\bibinfo{year}{2003}),
  \urlprefix\url{http://link.aps.org/doi/10.1103/RevModPhys.75.1101}.

\bibitem[{\citenamefont{Kivelson et~al.}(1997)\citenamefont{Kivelson, Lee,
  Krotov, and Gan}}]{klkgphhlr}
\bibinfo{author}{\bibfnamefont{S.~A.} \bibnamefont{Kivelson}},
  \bibinfo{author}{\bibfnamefont{D.-H.} \bibnamefont{Lee}},
  \bibinfo{author}{\bibfnamefont{Y.}~\bibnamefont{Krotov}}, \bibnamefont{and}
  \bibinfo{author}{\bibfnamefont{J.}~\bibnamefont{Gan}},
  \bibinfo{journal}{Phys. Rev. B} \textbf{\bibinfo{volume}{55}},
  \bibinfo{pages}{15552} (\bibinfo{year}{1997}),
  \urlprefix\url{http://link.aps.org/doi/10.1103/PhysRevB.55.15552}.

\bibitem[{\citenamefont{Rezayi and Haldane}(2000)}]{rzyhald2000}
\bibinfo{author}{\bibfnamefont{E.~H.} \bibnamefont{Rezayi}} \bibnamefont{and}
  \bibinfo{author}{\bibfnamefont{F.~D.~M.} \bibnamefont{Haldane}},
  \bibinfo{journal}{Phys. Rev. Lett.} \textbf{\bibinfo{volume}{84}},
  \bibinfo{pages}{4685} (\bibinfo{year}{2000}),
  \urlprefix\url{http://link.aps.org/doi/10.1103/PhysRevLett.84.4685}.

\bibitem[{\citenamefont{Geraedts et~al.}(2016)\citenamefont{Geraedts, Zaletel,
  Mong, Metlitski, Vishwanath, and Motrunich}}]{geraedts2016half}
\bibinfo{author}{\bibfnamefont{S.~D.} \bibnamefont{Geraedts}},
  \bibinfo{author}{\bibfnamefont{M.~P.} \bibnamefont{Zaletel}},
  \bibinfo{author}{\bibfnamefont{R.~S.} \bibnamefont{Mong}},
  \bibinfo{author}{\bibfnamefont{M.~A.} \bibnamefont{Metlitski}},
  \bibinfo{author}{\bibfnamefont{A.}~\bibnamefont{Vishwanath}},
  \bibnamefont{and} \bibinfo{author}{\bibfnamefont{O.~I.}
  \bibnamefont{Motrunich}}, \bibinfo{journal}{Science}
  \textbf{\bibinfo{volume}{352}}, \bibinfo{pages}{197} (\bibinfo{year}{2016}).

\bibitem[{\citenamefont{Son}(2015)}]{son2015composite}
\bibinfo{author}{\bibfnamefont{D.~T.} \bibnamefont{Son}},
  \bibinfo{journal}{Physical Review X} \textbf{\bibinfo{volume}{5}},
  \bibinfo{pages}{031027} (\bibinfo{year}{2015}).

\bibitem[{\citenamefont{Wang and Senthil}(2015)}]{wang2015dual}
\bibinfo{author}{\bibfnamefont{C.}~\bibnamefont{Wang}} \bibnamefont{and}
  \bibinfo{author}{\bibfnamefont{T.}~\bibnamefont{Senthil}},
  \bibinfo{journal}{Physical Review X} \textbf{\bibinfo{volume}{5}},
  \bibinfo{pages}{041031} (\bibinfo{year}{2015}).

\bibitem[{\citenamefont{Metlitski and
  Vishwanath}(2016)}]{metlitski2016particle}
\bibinfo{author}{\bibfnamefont{M.~A.} \bibnamefont{Metlitski}}
  \bibnamefont{and}
  \bibinfo{author}{\bibfnamefont{A.}~\bibnamefont{Vishwanath}},
  \bibinfo{journal}{Physical Review B} \textbf{\bibinfo{volume}{93}},
  \bibinfo{pages}{245151} (\bibinfo{year}{2016}).

\bibitem[{\citenamefont{Wang and Senthil}(2016{\natexlab{a}})}]{wang2016half}
\bibinfo{author}{\bibfnamefont{C.}~\bibnamefont{Wang}} \bibnamefont{and}
  \bibinfo{author}{\bibfnamefont{T.}~\bibnamefont{Senthil}},
  \bibinfo{journal}{Physical Review B} \textbf{\bibinfo{volume}{93}},
  \bibinfo{pages}{085110} (\bibinfo{year}{2016}{\natexlab{a}}).

\bibitem[{\citenamefont{Mross et~al.}(2016)\citenamefont{Mross, Alicea, and
  Motrunich}}]{mross2016explicit}
\bibinfo{author}{\bibfnamefont{D.~F.} \bibnamefont{Mross}},
  \bibinfo{author}{\bibfnamefont{J.}~\bibnamefont{Alicea}}, \bibnamefont{and}
  \bibinfo{author}{\bibfnamefont{O.~I.} \bibnamefont{Motrunich}},
  \bibinfo{journal}{Physical review letters} \textbf{\bibinfo{volume}{117}},
  \bibinfo{pages}{016802} (\bibinfo{year}{2016}).

\bibitem[{\citenamefont{Karch and Tong}(2016)}]{karch2016particle}
\bibinfo{author}{\bibfnamefont{A.}~\bibnamefont{Karch}} \bibnamefont{and}
  \bibinfo{author}{\bibfnamefont{D.}~\bibnamefont{Tong}},
  \bibinfo{journal}{Physical Review X} \textbf{\bibinfo{volume}{6}},
  \bibinfo{pages}{031043} (\bibinfo{year}{2016}).

\bibitem[{\citenamefont{Seiberg et~al.}(2016)\citenamefont{Seiberg, Senthil,
  Wang, and Witten}}]{seiberg2016duality}
\bibinfo{author}{\bibfnamefont{N.}~\bibnamefont{Seiberg}},
  \bibinfo{author}{\bibfnamefont{T.}~\bibnamefont{Senthil}},
  \bibinfo{author}{\bibfnamefont{C.}~\bibnamefont{Wang}}, \bibnamefont{and}
  \bibinfo{author}{\bibfnamefont{E.}~\bibnamefont{Witten}},
  \bibinfo{journal}{Annals of Physics} \textbf{\bibinfo{volume}{374}},
  \bibinfo{pages}{395} (\bibinfo{year}{2016}).

\bibitem[{\citenamefont{Wang et~al.}(2017)\citenamefont{Wang, Cooper, Halperin,
  and Stern}}]{wang2017particle}
\bibinfo{author}{\bibfnamefont{C.}~\bibnamefont{Wang}},
  \bibinfo{author}{\bibfnamefont{N.~R.} \bibnamefont{Cooper}},
  \bibinfo{author}{\bibfnamefont{B.~I.} \bibnamefont{Halperin}},
  \bibnamefont{and} \bibinfo{author}{\bibfnamefont{A.}~\bibnamefont{Stern}},
  \bibinfo{journal}{Physical Review X} \textbf{\bibinfo{volume}{7}},
  \bibinfo{pages}{031029} (\bibinfo{year}{2017}).

\bibitem[{\citenamefont{Kumar et~al.}(2019)\citenamefont{Kumar, Raghu, and
  Mulligan}}]{kumar2019composite}
\bibinfo{author}{\bibfnamefont{P.}~\bibnamefont{Kumar}},
  \bibinfo{author}{\bibfnamefont{S.}~\bibnamefont{Raghu}}, \bibnamefont{and}
  \bibinfo{author}{\bibfnamefont{M.}~\bibnamefont{Mulligan}},
  \bibinfo{journal}{Physical Review B} \textbf{\bibinfo{volume}{99}},
  \bibinfo{pages}{235114} (\bibinfo{year}{2019}).

\bibitem[{\citenamefont{Senthil et~al.}(2019)\citenamefont{Senthil, Son, Wang,
  and Xu}}]{senthil2019duality}
\bibinfo{author}{\bibfnamefont{T.}~\bibnamefont{Senthil}},
  \bibinfo{author}{\bibfnamefont{D.~T.} \bibnamefont{Son}},
  \bibinfo{author}{\bibfnamefont{C.}~\bibnamefont{Wang}}, \bibnamefont{and}
  \bibinfo{author}{\bibfnamefont{C.}~\bibnamefont{Xu}},
  \bibinfo{journal}{Physics Reports}  (\bibinfo{year}{2019}).

\bibitem[{\citenamefont{Shankar and Murthy}(1997)}]{rsgm97}
\bibinfo{author}{\bibfnamefont{R.}~\bibnamefont{Shankar}} \bibnamefont{and}
  \bibinfo{author}{\bibfnamefont{G.}~\bibnamefont{Murthy}},
  \bibinfo{journal}{Phys. Rev. Lett.} \textbf{\bibinfo{volume}{79}},
  \bibinfo{pages}{4437} (\bibinfo{year}{1997}),
  \urlprefix\url{http://link.aps.org/doi/10.1103/PhysRevLett.79.4437}.

\bibitem[{\citenamefont{Pasquier and Haldane}(1998)}]{Pasquier_1998}
\bibinfo{author}{\bibfnamefont{V.}~\bibnamefont{Pasquier}} \bibnamefont{and}
  \bibinfo{author}{\bibfnamefont{F.}~\bibnamefont{Haldane}},
  \bibinfo{journal}{Nuc. Phys. B} \textbf{\bibinfo{volume}{516}},
  \bibinfo{pages}{719–726} (\bibinfo{year}{1998}), ISSN
  \bibinfo{issn}{0550-3213},
  \urlprefix\url{http://dx.doi.org/10.1016/S0550-3213(98)00069-8}.

\bibitem[{\citenamefont{Read}(1998)}]{Read_1998}
\bibinfo{author}{\bibfnamefont{N.}~\bibnamefont{Read}}, \bibinfo{journal}{Phys.
  Rev. B} \textbf{\bibinfo{volume}{58}}, \bibinfo{pages}{16262–16290}
  (\bibinfo{year}{1998}), ISSN \bibinfo{issn}{1095-3795},
  \urlprefix\url{http://dx.doi.org/10.1103/PhysRevB.58.16262}.

\bibitem[{\citenamefont{Lee}(1998)}]{dhleephcf98}
\bibinfo{author}{\bibfnamefont{D.-H.} \bibnamefont{Lee}},
  \bibinfo{journal}{Phys. Rev. Lett.} \textbf{\bibinfo{volume}{80}},
  \bibinfo{pages}{4745} (\bibinfo{year}{1998}),
  \urlprefix\url{http://link.aps.org/doi/10.1103/PhysRevLett.80.4745}.

\bibitem[{\citenamefont{Stern et~al.}(1999)\citenamefont{Stern, Halperin, von
  Oppen, and Simon}}]{sternetal99}
\bibinfo{author}{\bibfnamefont{A.}~\bibnamefont{Stern}},
  \bibinfo{author}{\bibfnamefont{B.~I.} \bibnamefont{Halperin}},
  \bibinfo{author}{\bibfnamefont{F.}~\bibnamefont{von Oppen}},
  \bibnamefont{and} \bibinfo{author}{\bibfnamefont{S.~H.} \bibnamefont{Simon}},
  \bibinfo{journal}{Phys. Rev. B} \textbf{\bibinfo{volume}{59}},
  \bibinfo{pages}{12547} (\bibinfo{year}{1999}),
  \urlprefix\url{http://link.aps.org/doi/10.1103/PhysRevB.59.12547}.

\bibitem[{\citenamefont{Cooper et~al.}(2001)\citenamefont{Cooper, Wilkin, and
  Gunn}}]{cooper2001quantum}
\bibinfo{author}{\bibfnamefont{N.~R.} \bibnamefont{Cooper}},
  \bibinfo{author}{\bibfnamefont{N.~K.} \bibnamefont{Wilkin}},
  \bibnamefont{and} \bibinfo{author}{\bibfnamefont{J.}~\bibnamefont{Gunn}},
  \bibinfo{journal}{Physical review letters} \textbf{\bibinfo{volume}{87}},
  \bibinfo{pages}{120405} (\bibinfo{year}{2001}).

\bibitem[{\citenamefont{Alicea et~al.}(2005)\citenamefont{Alicea, Motrunich,
  Hermele, and Fisher}}]{avl1}
\bibinfo{author}{\bibfnamefont{J.}~\bibnamefont{Alicea}},
  \bibinfo{author}{\bibfnamefont{O.~I.} \bibnamefont{Motrunich}},
  \bibinfo{author}{\bibfnamefont{M.}~\bibnamefont{Hermele}}, \bibnamefont{and}
  \bibinfo{author}{\bibfnamefont{M.~P.~A.} \bibnamefont{Fisher}},
  \bibinfo{journal}{Phys. Rev. B} \textbf{\bibinfo{volume}{72}},
  \bibinfo{pages}{064407} (\bibinfo{year}{2005}),
  \urlprefix\url{http://link.aps.org/doi/10.1103/PhysRevB.72.064407}.

\bibitem[{\citenamefont{Wang and
  Senthil}(2016{\natexlab{b}})}]{wang2016composite}
\bibinfo{author}{\bibfnamefont{C.}~\bibnamefont{Wang}} \bibnamefont{and}
  \bibinfo{author}{\bibfnamefont{T.}~\bibnamefont{Senthil}},
  \bibinfo{journal}{Physical Review B} \textbf{\bibinfo{volume}{94}},
  \bibinfo{pages}{245107} (\bibinfo{year}{2016}{\natexlab{b}}).

\bibitem[{\citenamefont{Seiberg and Witten}(1999)}]{Seiberg_1999}
\bibinfo{author}{\bibfnamefont{N.}~\bibnamefont{Seiberg}} \bibnamefont{and}
  \bibinfo{author}{\bibfnamefont{E.}~\bibnamefont{Witten}},
  \bibinfo{journal}{Journal of High Energy Physics}
  \textbf{\bibinfo{volume}{1999}}, \bibinfo{pages}{032} (\bibinfo{year}{1999}),
  \urlprefix\url{https://doi.org/10.1088%2F1126-6708%2F1999%2F09%2F032}.

\bibitem[{\citenamefont{Snyder}(1947)}]{snyder}
\bibinfo{author}{\bibfnamefont{H.~S.} \bibnamefont{Snyder}},
  \bibinfo{journal}{Phys. Rev.} \textbf{\bibinfo{volume}{71}},
  \bibinfo{pages}{38} (\bibinfo{year}{1947}),
  \urlprefix\url{https://link.aps.org/doi/10.1103/PhysRev.71.38}.

\bibitem[{\citenamefont{Douglas and
  Nekrasov}(2001)}]{douglas2001noncommutative}
\bibinfo{author}{\bibfnamefont{M.~R.} \bibnamefont{Douglas}} \bibnamefont{and}
  \bibinfo{author}{\bibfnamefont{N.~A.} \bibnamefont{Nekrasov}},
  \bibinfo{journal}{Reviews of Modern Physics} \textbf{\bibinfo{volume}{73}},
  \bibinfo{pages}{977} (\bibinfo{year}{2001}).

\bibitem[{\citenamefont{Szabo}(2003)}]{szabo2003quantum}
\bibinfo{author}{\bibfnamefont{R.~J.} \bibnamefont{Szabo}},
  \bibinfo{journal}{Physics Reports} \textbf{\bibinfo{volume}{378}},
  \bibinfo{pages}{207} (\bibinfo{year}{2003}).

\bibitem[{\citenamefont{Bellissard et~al.}(1994)\citenamefont{Bellissard, van
  Elst, and Schulz-Baldes}}]{bellissard1994noncommutative}
\bibinfo{author}{\bibfnamefont{J.}~\bibnamefont{Bellissard}},
  \bibinfo{author}{\bibfnamefont{A.}~\bibnamefont{van Elst}}, \bibnamefont{and}
  \bibinfo{author}{\bibfnamefont{H.}~\bibnamefont{Schulz-Baldes}},
  \bibinfo{journal}{Journal of Mathematical Physics}
  \textbf{\bibinfo{volume}{35}}, \bibinfo{pages}{5373} (\bibinfo{year}{1994}).

\bibitem[{\citenamefont{Susskind}(2001)}]{susskind2001quantum}
\bibinfo{author}{\bibfnamefont{L.}~\bibnamefont{Susskind}},
  \bibinfo{journal}{arXiv preprint hep-th/0101029}  (\bibinfo{year}{2001}).

\bibitem[{\citenamefont{Polychronakos}(2001)}]{polychronakos2001quantum}
\bibinfo{author}{\bibfnamefont{A.~P.} \bibnamefont{Polychronakos}},
  \bibinfo{journal}{Journal of High Energy Physics}
  \textbf{\bibinfo{volume}{2001}}, \bibinfo{pages}{011} (\bibinfo{year}{2001}).

\bibitem[{\citenamefont{Haldane}(2011)}]{haldane2011geometrical}
\bibinfo{author}{\bibfnamefont{F.}~\bibnamefont{Haldane}},
  \bibinfo{journal}{Physical review letters} \textbf{\bibinfo{volume}{107}},
  \bibinfo{pages}{116801} (\bibinfo{year}{2011}).

\bibitem[{\citenamefont{Girvin et~al.}(1986)\citenamefont{Girvin, MacDonald,
  and Platzman}}]{GMP}
\bibinfo{author}{\bibfnamefont{S.~M.} \bibnamefont{Girvin}},
  \bibinfo{author}{\bibfnamefont{A.~H.} \bibnamefont{MacDonald}},
  \bibnamefont{and} \bibinfo{author}{\bibfnamefont{P.~M.}
  \bibnamefont{Platzman}}, \bibinfo{journal}{Phys. Rev. B}
  \textbf{\bibinfo{volume}{33}}, \bibinfo{pages}{2481} (\bibinfo{year}{1986}),
  \urlprefix\url{https://link.aps.org/doi/10.1103/PhysRevB.33.2481}.

\bibitem[{\citenamefont{Liu}(2001)}]{liu2001trek}
\bibinfo{author}{\bibfnamefont{H.}~\bibnamefont{Liu}},
  \bibinfo{journal}{Nuclear Physics B} \textbf{\bibinfo{volume}{614}},
  \bibinfo{pages}{305} (\bibinfo{year}{2001}).

\bibitem[{\citenamefont{Okawa and Ooguri}(2001)}]{okawa2001exact}
\bibinfo{author}{\bibfnamefont{Y.}~\bibnamefont{Okawa}} \bibnamefont{and}
  \bibinfo{author}{\bibfnamefont{H.}~\bibnamefont{Ooguri}},
  \bibinfo{journal}{Physical Review D} \textbf{\bibinfo{volume}{64}},
  \bibinfo{pages}{046009} (\bibinfo{year}{2001}).

\bibitem[{\citenamefont{Mukhi and Suryanarayana}(2001)}]{mukhi2001gauge}
\bibinfo{author}{\bibfnamefont{S.}~\bibnamefont{Mukhi}} \bibnamefont{and}
  \bibinfo{author}{\bibfnamefont{N.~V.} \bibnamefont{Suryanarayana}},
  \bibinfo{journal}{Journal of High Energy Physics}
  \textbf{\bibinfo{volume}{2001}}, \bibinfo{pages}{023} (\bibinfo{year}{2001}).

\bibitem[{\citenamefont{Liu and Michelson}(2001)}]{liu2001ramond}
\bibinfo{author}{\bibfnamefont{H.}~\bibnamefont{Liu}} \bibnamefont{and}
  \bibinfo{author}{\bibfnamefont{J.}~\bibnamefont{Michelson}},
  \bibinfo{journal}{Physics Letters B} \textbf{\bibinfo{volume}{518}},
  \bibinfo{pages}{143} (\bibinfo{year}{2001}).

\bibitem[{\citenamefont{Jackiw et~al.}(2002)\citenamefont{Jackiw, Pi, and
  Polychronakos}}]{jackiw2002noncommuting}
\bibinfo{author}{\bibfnamefont{R.}~\bibnamefont{Jackiw}},
  \bibinfo{author}{\bibfnamefont{S.-Y.} \bibnamefont{Pi}}, \bibnamefont{and}
  \bibinfo{author}{\bibfnamefont{A.}~\bibnamefont{Polychronakos}},
  \bibinfo{journal}{Annals of Physics} \textbf{\bibinfo{volume}{301}},
  \bibinfo{pages}{157} (\bibinfo{year}{2002}).

\bibitem[{\citenamefont{Wu and Jain}(2015)}]{wu2015emergent}
\bibinfo{author}{\bibfnamefont{Y.-H.} \bibnamefont{Wu}} \bibnamefont{and}
  \bibinfo{author}{\bibfnamefont{J.~K.} \bibnamefont{Jain}},
  \bibinfo{journal}{Physical Review A} \textbf{\bibinfo{volume}{91}},
  \bibinfo{pages}{063623} (\bibinfo{year}{2015}).

\bibitem[{\citenamefont{Geraedts et~al.}(2017)\citenamefont{Geraedts, Repellin,
  Wang, Mong, Senthil, and Regnault}}]{geraedts2017emergent}
\bibinfo{author}{\bibfnamefont{S.~D.} \bibnamefont{Geraedts}},
  \bibinfo{author}{\bibfnamefont{C.}~\bibnamefont{Repellin}},
  \bibinfo{author}{\bibfnamefont{C.}~\bibnamefont{Wang}},
  \bibinfo{author}{\bibfnamefont{R.~S.} \bibnamefont{Mong}},
  \bibinfo{author}{\bibfnamefont{T.}~\bibnamefont{Senthil}}, \bibnamefont{and}
  \bibinfo{author}{\bibfnamefont{N.}~\bibnamefont{Regnault}},
  \bibinfo{journal}{Physical Review B} \textbf{\bibinfo{volume}{96}},
  \bibinfo{pages}{075148} (\bibinfo{year}{2017}).

\bibitem[{\citenamefont{Read}(1994)}]{read94}
\bibinfo{author}{\bibfnamefont{N.}~\bibnamefont{Read}},
  \bibinfo{journal}{Semiconductor Science and Technology}
  \textbf{\bibinfo{volume}{9}}, \bibinfo{pages}{1859} (\bibinfo{year}{1994}),
  \urlprefix\url{http://stacks.iop.org/0268-1242/9/i=11S/a=002}.

\bibitem[{\citenamefont{Murthy and Shankar}(2016)}]{murthy2016nu}
\bibinfo{author}{\bibfnamefont{G.}~\bibnamefont{Murthy}} \bibnamefont{and}
  \bibinfo{author}{\bibfnamefont{R.}~\bibnamefont{Shankar}},
  \bibinfo{journal}{Physical Review B} \textbf{\bibinfo{volume}{93}},
  \bibinfo{pages}{085405} (\bibinfo{year}{2016}).

\bibitem[{\citenamefont{Green}(2002)}]{green2002strongly}
\bibinfo{author}{\bibfnamefont{D.}~\bibnamefont{Green}},
  \bibinfo{journal}{arXiv preprint cond-mat/0202455}  (\bibinfo{year}{2002}).

\end{thebibliography}
\end{document}